\documentclass[twocolumn, prl, aps, superscriptaddress, longbibliography,showpacs,amsmath,amssymb,floatfix
]{revtex4-1}

\usepackage{graphicx}
\usepackage{dcolumn}
\usepackage{bm}
\usepackage{graphicx}                                               
\usepackage{amssymb}

\usepackage{amsmath}
\usepackage{epsfig}
\usepackage{xcolor}
\usepackage{tabu}
\usepackage{mathtools}
\usepackage[colorlinks,linkcolor=blue,anchorcolor=blue,citecolor=blue,urlcolor=blue]{hyperref}
\usepackage{physics}
\usepackage{float}
\usepackage{diagbox}
\usepackage{inputenc}

\begin{document}

\title{Fate of localization in coupled free chain and disordered chain}
\author{Xiaoshui Lin}
\affiliation{CAS Key Laboratory of Quantum Information, University of Science and Technology of China, Hefei, 230026, China}
\author{Ming Gong}
\email{gongm@ustc.edu.cn}
\affiliation{CAS Key Laboratory of Quantum Information, University of Science and Technology of China, Hefei, 230026, China}
\affiliation{Synergetic Innovation Center of Quantum Information and Quantum Physics, University of Science and Technology of China, Hefei, Anhui 230026, China}
\affiliation{Hefei National Laboratory, University of Science and Technology of China, Hefei 230088, China}

\date{\today}

\begin{abstract}
It has been widely believed that almost all states in one-dimensional (1d) disordered systems with short-range hopping and uncorrelated random potential are localized. 
Here, we consider the fate of these localized states by coupling between a disordered chain (with localized states) and a free chain (with extended states), showing that states in the overlapped and un-overlapped regimes exhibit totally different localization behaviors, which is not a phase transition process. 
In particular, while states in the overlapped regime are localized by resonant coupling, in the un-overlapped regime of the free chain, significant suppression of the localization with a prefactor of $\xi^{-1} \propto t_v^4/\Delta^4$ appeared, where $t_v$ is the inter-chain coupling strength and $\Delta$ is the energy shift between them. 
This system may exhibit localization lengths that are comparable with the system size even when the potential in the disordered chain is strong.
We confirm these results using the transfer matrix method and sparse matrix method for systems $L \sim 10^6 - 10^9$.
These findings extend our understanding of localization in low-dimensional disordered systems and provide a concrete example, 
which may call for much more advanced numerical methods in high-dimensional models. 
\end{abstract}

\maketitle

Anderson localization (AL), which describes the phenomenon that the disorder totally suppresses the diffusion of the system, has attracted a great deal of attention for many decades \citep{anderson_absence_1958, Lee1985Disordered, Evers2008Anderson, Lagendijk2009Fifty, Roati2008Anderson, Segev2013Anderson}.
It has been found that the spatial dimension plays an essential role in AL \citep{Mott1961Theory, Abrahams1979Scaling, Ishii1973Localization, Thouless1972Relation, Thouless1973Localization}.
In the one-dimensional (1d) tight-binding model with random potential, the localization length is given by \citep{Thouless1981Conductivity, Heinrichs2002Localization}
\begin{equation}
\xi_0^{-1}(E) = {v^2 \over 8 t^2 - 2 E^2} ={V^2 \over 96 t^2 - 24 E^2},
\label{eq-thouless-expression}
\end{equation} 
where $v^2=\langle v_i^2\rangle$ is the variance of the potential $v_i \in U[-V/2, V/2]$, with $V$ being the disorder strength, $t$ is the hopping strength between the neighboring sites and $E$ is the eigenvalue (see Eq. 14 in Ref. \citep{Thouless1981Conductivity} for more details ). 
When the system size $L$ is much larger than $\xi_0$, $L \gg \xi_0$, localization of wave functions can be observed and the system is in the localized phase without conductance. 
In the presence of weak disorder, $V \ll t$, we have $|E| < 2t$, thus all states should be localized with $\xi_{0}^{-1}>0$. 
For example, when $V \sim 0.1t$, $\xi_0 \sim 10^4$, which can be easily confirmed by numerical simulation.
The fate of the states in 1d systems will be changed fundamentally with incommensurate potentials \citep{Biddle2009Localization, Biddle2010Predicted, Ganeshan2015Nearest, Wang2020one-dimensional}, long-range correlated disorders \citep{Izrailev1999Localization,  Moura1998Delocalization, Dietz2011Microwave,Delande2014Mobility, Johnston1986Localization} and many-body interactions \citep{Bordia2016Coupling, Thiery2018Many, Chiew2023Stability, Leonard2023Probing, Hyunsoo2023Many-Body}. 

\begin{figure}[htbp]
\centering
\includegraphics[width=0.45\textwidth]{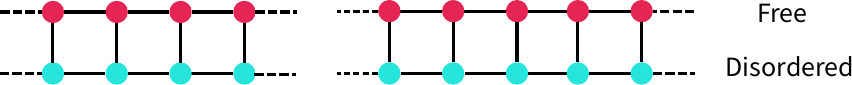}
\caption{The realization of the coupled disordered model. The free chain does not have random potential, thus all states are extended; yet in the disordered chain with random potential, all states are localized. 
The coupling between them is the major concern of this work.}
\label{fig-model}
\end{figure}

While the physics of disordered models have been widely discussed \citep{Ishii1973Localization, Brouwer1998Delocalization, Heinrichs2002Localization, Evers2008Anderson}, the fate of localization by the coupling of extended and localized states are much lesser investigated. 
We are interested in this issue for the reason of the dilemma that (I) the random uncorrelated potential induces localization for the extended states in 1d systems \citep{Abrahams1979Scaling}; 
and (II) the hybridization between localized and extended states may lead to delocalization  \citep{Mott1987Mobility}.
The interplay between these two mechanisms may lead to different physics. 
In this work, we propose a coupled disordered model (Fig. \ref{fig-model}) to address this problem.
Our model is constructed from one free chain (with all states extended) and one disordered chain (with all states localized). Two major conclusions have been established: 
(1) While all states exhibit localization in the presence of inter-chain coupling, their localization length exhibits a distinct difference in the overlapped spectra and 
un-overlapped spectra, which is not a phase transition process; (2) The localization length for states in the un-overlapped regime 
of the free chain is significantly suppressed given by 
\begin{equation}
\xi^{-1}(E) \simeq {t_\text{v}^4 V^2 \over (96 t^2 - 24 (E - \Delta)^2) \Delta^4} = {t_\text{v}^4 \over \Delta^4} \xi_0^{-1}(E - \Delta), 
\label{eq-loc-length}
\end{equation}
in the limit when $\Delta \gg |V|$. 
Here $t_\text{v}$ is the inter-chain coupling, and $\Delta$ is the energy shift between the free and disordered chains. 
Thus localization is greatly suppressed when $t_\text{v} \ll \Delta$.
For instance, when $t_\text{v} = 0.1t$ and $\Delta =10t$, the localization length can be suppressed by eight orders of magnitude.
We examine the above conclusions using the transfer matrix method and the sparse matrix method with system sizes  $L\sim 10^6-10^9$.
Our results show that the inter-chain coupling, disorder strength, and energy shift are the three major factors influencing the localization length of states in the un-overlapped regime.
In the regime when $\xi \gtrsim L$, we can understand the localization of wave functions with the following general theorem based on the large amount of researches; see review articles \citep{Ishii1973Localization, Beenakker1997Random-matrix, Belitz1994Anderson, Evers2008Anderson}. 

\begin{figure}[htbp]
\centering
\includegraphics[width=0.45\textwidth]{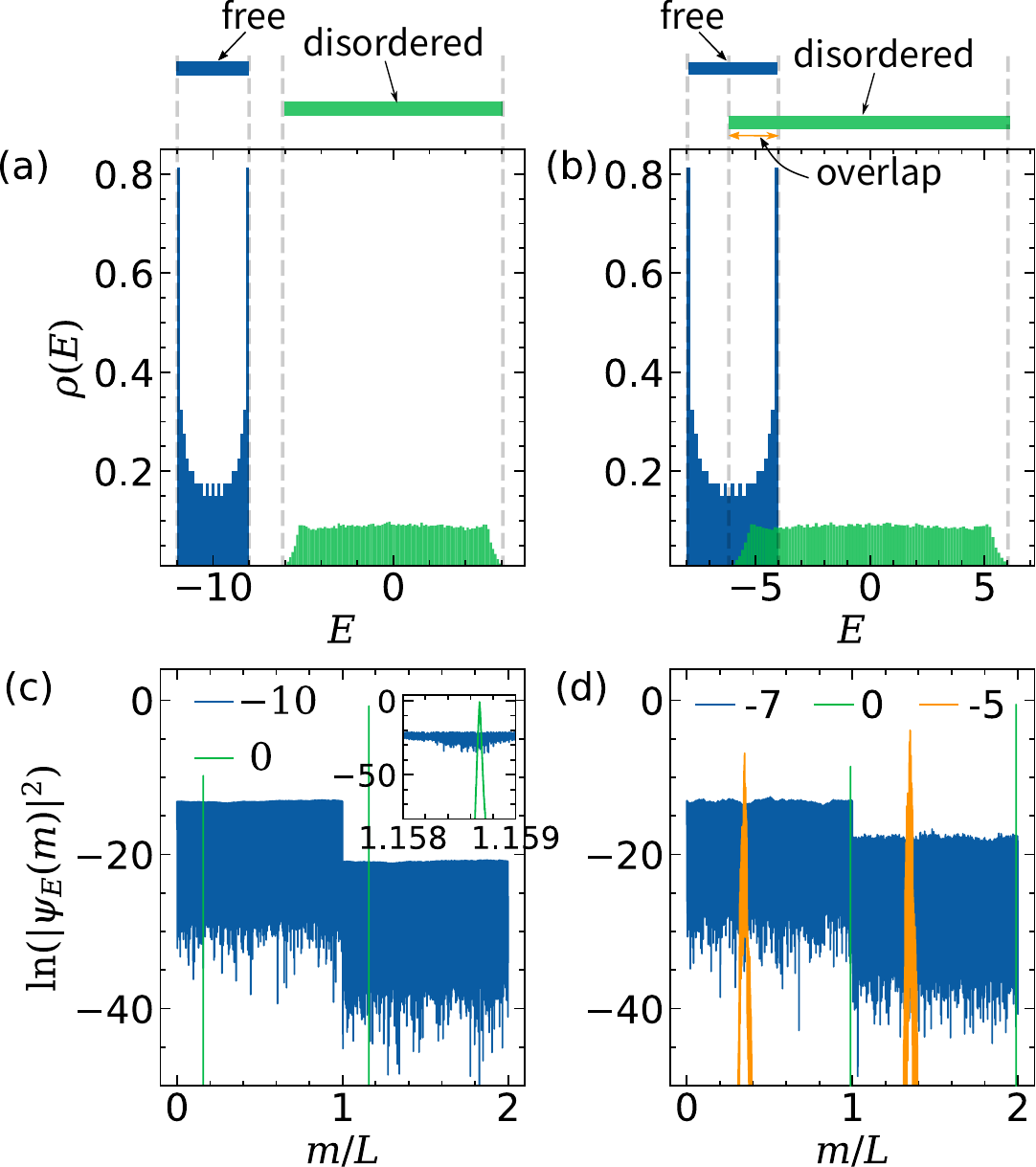}
\caption{(a) (b) The averaged density of states $\rho(E)$ for the free and disordered chains with $t_\text{v} = 0$. Other parameters are (a) $\Delta = -10$, 
$V = 10$; and (b) $\Delta=-6$ and $V = 10$.
(c) (d) Logarithm of the wave function amplitude with different energy against the lattice index $m$ for the coupled disordered model ($t_\text{v} = 0.1$).
The index $m<L$ ($m\geq L$) is for Hilbert space of $H_0$ ($H_1$).
Inset of (c) shows the detailed wave function around its localized center. In panel (a) and (b), $L=10^3$; (c) and (d) $L=10^6$.}
\label{fig-wave-function}
\end{figure}

\textit{Theorem}: In 1d disordered systems with short-range hopping and uncorrelated random potential, almost all states are localized in the thermodynamic limit ($L \rightarrow \infty$). 

The exact proof of this theorem is still a great challenge at the present stage; 
however, it can be understood intuitively from the observation by Mott \textit{et al} \citep{Mott1961Theory}, the argument by Thouless \citep{thouless1974electrons}, and the scaling argument by Abrahams \textit{et al}, in which the $\beta$ function is always negative \citep{Abrahams1979Scaling}. 
This theorem is also addressed by the celebrated Dorokhov-Mello-Pereyra-Kumar equation \citep{Beenakker1997Random-matrix} and the non-linear sigma model
\citep{efetov1983supersymmetry}.
Here, not all states are localized because, in the 1d model with off-diagonal random potential, the state with $E = 0$ is extended while all the other states are localized \citep{George1976Extended}.
The mathematicians have great interest in this problem and have proved this theorem with random potentials \citep{Carmona1987Anderson, Kunz1980Sur, Frohlich1983Absence, Pastur1980Spectral}, showing of no continuous spectra for extended states. 
This theorem will play a determinative role for localization when $\xi \gtrsim L$, which is beyond the capability of the numerical simulations. 

\textit{Physical model and methods}:
We consider the following coupled disordered model (see Fig. \ref{fig-model})
\begin{equation}
    H = H_0 + H_1 + \sum_{m,\sigma} t_\text{v} a_{m,\sigma}^{\dagger}a_{m,\bar{\sigma}},
    \label{eq-model}
\end{equation}
where $H_\sigma = \sum_{m} t_{\sigma} a_{m,\sigma}^{\dagger}a_{m+1,\sigma} + \mathrm{h.c.} + V_{m,\sigma}a_{m,\sigma}^{\dagger}a_{m,\sigma}$, with $H_0$ for the free chain and $H_1$ for the disordered chain. 
Here $V_{m, 0} = \Delta$ is the energy shift of the free chain $H_0$ and 
$V_{m, 1} \in U[- V/2, V/2]$ is the random potential in the disordered chain $H_1$, and $\bar{\sigma} = 1-\sigma \in \{ 0, 1\}$.
When $t_\text{v} = 0$, this model is reduced to a free chain with all states extended, and a disordered chain with all states localized.
By changing $\Delta$, the energy spectra of these two chains can be un-overlapped (Fig. \ref{fig-wave-function} (a)) or overlapped (Fig. \ref{fig-wave-function} (b)).
By the theorem, all states should be localized in the presence of inter-chain coupling (with $t_v \ne 0$). The fundamental question is what are the quantitative differences between the states in the overlapped regime and the un-overlapped regime of the coupled model during localization? 

We apply the transfer matrix and sparse matrix methods, whose available size is $L \sim 10^{6} - 10^{9}$, to understand the localization of wave functions in these two regimes.
In the transfer matrix method \citep{Hoffmanbook}, the Lyapunov exponent $\gamma(E) = \xi(E)^{-1}$ is defined as the smallest positive eigenvalue of the matrix 
\begin{equation}
    \Gamma(E) = \lim_{L\rightarrow \infty} \frac{1}{2L}\ln(T_1^{\dagger} \dots T_L^{\dagger}T_L\dots T_1),
    \label{eq-gammaE}
\end{equation}
where $T_i$ is the transfer matrix at the $i$-th site. 
From the Oseledets ergodic theorem \citep{Oseledets1968Multiplicative, Raghunathan1979Proof}, the above multiplication of transfer matrices is converged when $L \rightarrow \infty$.
When $\gamma(E) \neq 0$, the state with eigenvalue $E$ is localized. 
In the sparse matrix method, we use the shift-invert method \citep{Pietracaprina2018Shift, Luitz2015Many-body} to obtain about $N_E = 20$ eigenstates $|\psi_{E_i}\rangle$ with eigenvalues $E_i$ around a given $E$ and define the averaged inverse participation ratio (IPR) as   \citep{Evers2008Anderson} 
\begin{equation}
    \langle \text{IPR}\rangle_E = {1\over N_E} \sum_{i=1}^{N_E} \sum_{m=0}^{2L-1}|\psi_{E_i}(m)|^4. 
    \label{eq-iprE}
\end{equation}
For the extended state, $\langle \text{IPR}\rangle_E \propto L^{-1}$ and for the localized state, $\langle \text{IPR}\rangle_E$ is finite.
Furthermore, we can define the fractal dimension $\tau_2(E, L) = -\ln(\langle \text{IPR}\rangle_E)/\ln(L)$ and its limit $\tau_2(E) = \lim_{L\rightarrow\infty} \tau_2(E, L)$.
We have $\tau_2(E) = 0$ for localized states, $\tau_2(E) = 1$ for extended states, and $0<\tau_2(E)<1$ for critical states, respectively \citep{Hiramoto1989Scaling, Evers2008Anderson, Lin2022General}.
We note that the IPR should be proportional to the Lyapunov exponent $\gamma(E)$ for an exponentially localized state $\psi_m \sim e^{-|m|/\xi}$, with $\text{IPR} \propto \xi^{-1} = \gamma$, in the limit $L \gg \xi$. 

\textit{Physics in the overlapped and un-overlapped regimes}:
Although all states are expected to be localized in our model, the effect of inter-chain coupling in the overlapped regime and un-overlapped regime should be different, leading to distinct localization behavior.
We consider two different cases, which are shown in Fig. \ref{fig-wave-function} (a) and (b). 
In the first case, the spectra in the two chains are un-overlapped to avoid the resonant coupling between the extended and localized states; while in the second case, resonant coupling is induced in their overlapped regime. Furthermore, we present their typical wave functions in these regimes in Fig. \ref{fig-wave-function} (c) and (d) with $t_\text{v}=0.1$.
The results show that the wave functions in the un-overlapped regime of the disordered chain are exponentially localized with localization length around unity (see the state with $E=0$ in Fig. \ref{fig-wave-function} (c) and (d)).
The wave functions in the overlapped regime are also exponentially localized, however, with localization length $\xi \sim 0.05L$, which is much larger than the localized states in the un-overlapped regime of the disordered chain.
Strikingly, the wave functions in the un-overlapped regime of the free chain are extended even when the system size $L=10^6$. 
Similar features can be found when $L$ is increased to $L \sim 10^9$ for smaller $t_\text{v}$. This seems to contradict with the general theorem. This dilemma is the major concern of this work. 

\begin{figure}[htbp]
\centering
\includegraphics[width=0.45\textwidth]{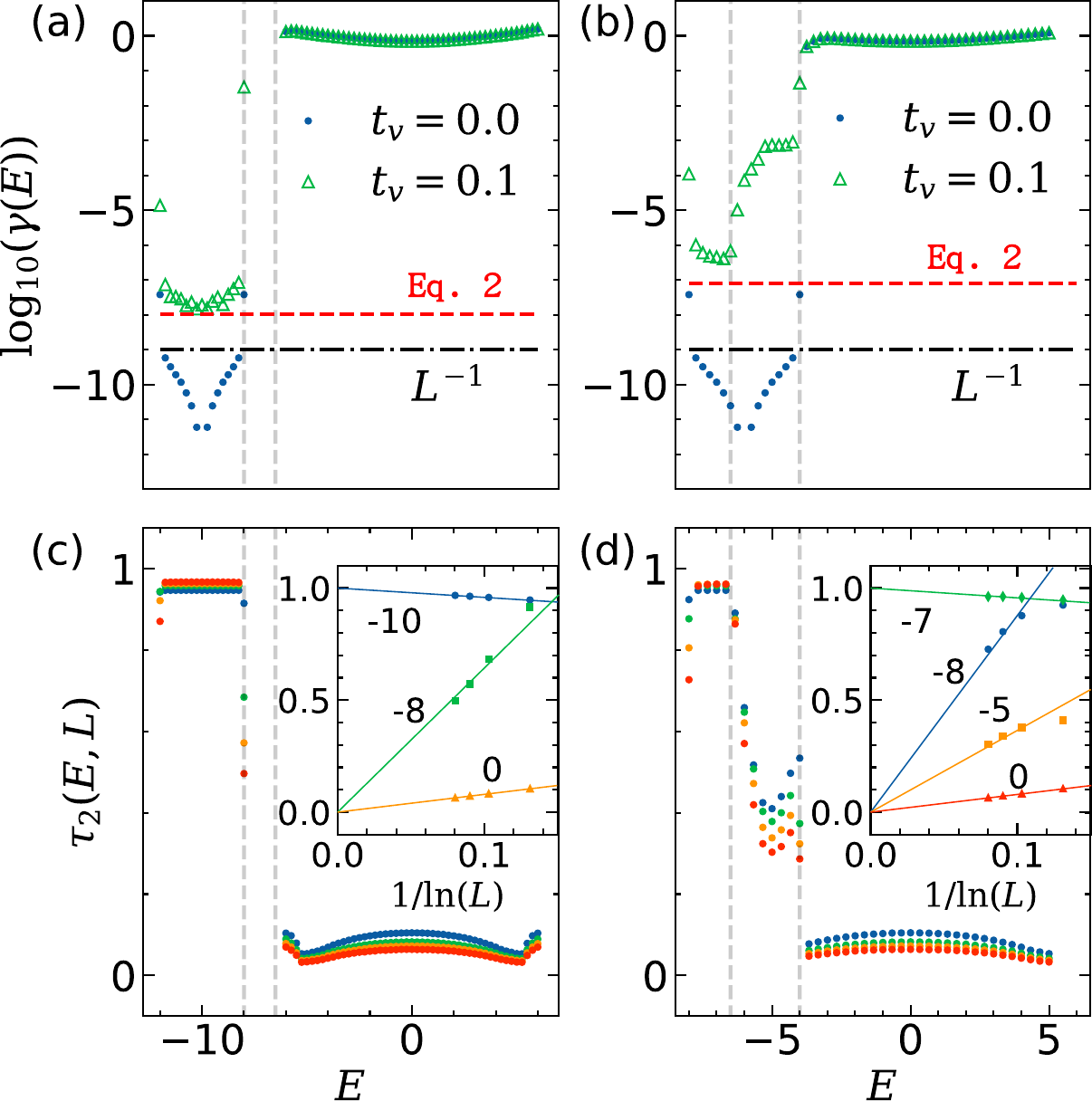}
\caption{
(a) (b) The Lyapunov exponent $\gamma(E)$ against the energy $E$ for $t_\text{v}=0$ and $t_\text{v}=0.1$, with system size $L = 10^9$. The shift is (a) $\Delta = -10$; (b) $\Delta=-6$. 
The vertical dashed lines in (a) denote $E=-8$ and $E=-6.4$, and in (b) denote $E=-6.4$ and $E=-4$.
The red dashed lines are estimated by Eq. \ref{eq-loc-length}.
(c) (d) The fractal dimension $\tau_2(E, L)$ versus energy $E$ for $t_\text{v}=0.1$.
The system size is $L=2^{11}$ (blue); $L=2^{14}$ (green); $L=2^{16}$ (yellow); $L=2^{18}$ (red). 
The inset present the results $\tau_2(E, L)$ against $1/\ln(L)$ for different energy $E$.
}
\label{fig-localization-length}
\end{figure}

Next, we investigate the asymptotic behavior of the wave function using the transfer matrix method.
In Fig. \ref{fig-localization-length} (a) and (b), we present the Lyapunov exponent $\gamma(E)$ against the energy $E$ for a system with size $L=10^9$.
When $t_\text{v} = 0$, the states in the free chain are extended and the states in the disordered chain are localized.
When $t_\text{v} = 0.1$, all states become localized with $\gamma(E)>L^{-1}$.
However, there are three distinct energy regimes for $\gamma(E)$, corresponding to the overlapped and un-overlapped regimes.
In the overlapped regime, we have $\gamma(E) \sim 10^{-4}$, while in the un-overlapped regime, we have $\gamma(E) \sim 10^{0}$ (disordered chain) or $\gamma(E) \sim 10^{-7}$ (free chain). 
These distinct behaviors are unique features of the coupled disorder model, which should not be regarded as some kind of phase transition between extended and localized states; see below. 

We further characterize the three energy regimes using IPR and fractal dimension $\tau_2(E, L)$, with results presented in Fig. \ref{fig-localization-length} (c) and (d).
It is found that $\tau_2(E, L) \rightarrow 0$ for states in the overlapped regime and un-overlapped regime of the disordered chain, indicating localization. 
However, $\tau_2(E, L) \rightarrow 1$ for states in the un-overlapped regime of the free chain, contradicting the previous theorem at first sight.
This contradiction is actually a finite-size effect since $L=2^{18} \sim 10^5 <\xi\sim 10^7$.
Thus, it is expected that all these states will be localized from the general theorem in the thermodynamic limit ($L \gg \xi$, or equivalently, $L\gamma(E) \gg 1$).
This also clarifies the previous disagreement presented in Fig. \ref{fig-wave-function}.

\textit{Origin of the suppressed localization length}:
The above results raise some fundamental questions that need to be addressed much more carefully. 
In the overlapped regime, the localized states and extended states are coupled through resonant coupling because their energy is close to each other.
From perturbation theory \cite{anderson_absence_1958}, all the higher-order terms will become important, leading to significant modification of the wave functions for localization.
In the un-overlapped regime of the free chain, wave function localization is greatly suppressed by a different mechanism.

\begin{figure}[htbp]
\centering
\includegraphics[width=0.45\textwidth]{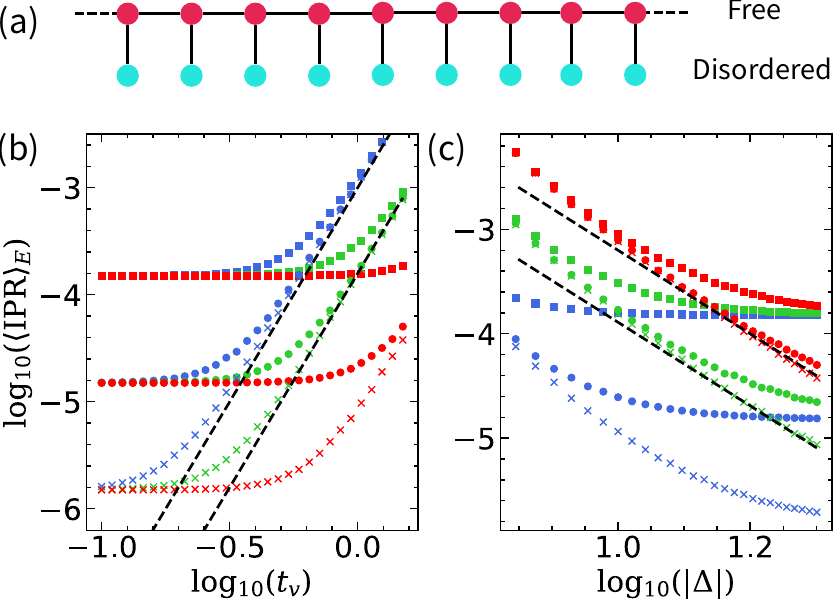}
\caption{
(a) The schematic of the model in Eq. \ref{eq-couple-full-localized}. 
The free chain $H_0$ (red) is coupled to a disordered chain $H_1$ (blue) with $t_1=0$.
(b) $\log_{10}(\text{IPR})$ against $\log_{10}(t_\text{v})$ for different energy shift $\Delta$ and lattice size with $E=\Delta$. The colors represent $\Delta =-20$ (red); $\Delta=-10$ (green); $\Delta=-7$ (blue).
The system size is $L=10^4$ (square), $10^5$ (circular), and $10^6$ (cross).
The two black dashed lines are linear fitting with $\log_{10}(\langle \text{IPR} \rangle_E) \sim  4\log_{10}(t_\text{v})$.
(c) $\log_{10}(\text{IPR})$ against $\log_{10}(|\Delta|)$ for different $t_\text{v}$. The meaning of symbols is the same as (b). 
The colors represent $t_\text{v} =1.5$ (red); $t_\text{v} = 1.0$ (green); and 
 $t_\text{v} =0.5$ (blue), and the black dashed lines denote $\log_{10}(\langle \text{IPR} \rangle_E) \sim  -4\log_{10}(|\Delta|)$.}
\label{fig-full-localized-chain}
\end{figure}

To this end, we first consider the localization in the following minimal model 
\begin{eqnarray}
    H = H_0 + (t_\text{v} a_{m,0}^{\dagger}a_{m,1} + \mathrm{h.c.}) + V_{m,1} a_{m,1}^{\dagger}a_{m,1}.
    \label{eq-minimal-model}
\end{eqnarray}
As compared with Eq. \ref{eq-model}, here we set $t_1 = 0$ and $t_0 = t = 1$ (see Fig. \ref{fig-full-localized-chain} (a)). 
When $t_\text{v}=0$, the eigenstates of component $\sigma=1$ are fully localized at one site with eigenvalue distributed in the interval $[-V/2, V/2]$.
On the other hand, the eigenstates of component $\sigma=0$ are $\psi(m) \propto e^{i k m}$ with energy spectra in $[\Delta - 2t, \Delta + 2t]$.
Thus, the energy spectra of each component are well separated when $|\Delta| > |V/2| + 2t$. We focus on the physics in this regime (with $t_\text{v} \neq 0$) for the suppressed localization. 

The central problem is to verify the major result of Eq. \ref{eq-loc-length}.
We employ the sparse matrix method to examine the physic in Eq. \ref{eq-minimal-model}, and the results for $E = \Delta$ are presented in Fig. \ref{fig-full-localized-chain}.
It is found that $\langle \text{IPR} \rangle_E \propto t_\text{v}^4$ with intermediate $t_\text{v}$ with $\xi < L$.
In the small inter-chain coupling limit, the IPR will be saturated to $L^{-1}$ when $\xi \gtrsim L$ due to the finite size effect.
Thus the relation of $\text{IPR} \propto t_\text{v}^4$ always holds in the thermodynamical limit.
In Fig. \ref{fig-full-localized-chain} (c), we also examine the dependence of IPR as a function of energy shift $\Delta$, finding that $\text{IPR} \propto \Delta^{-4}$ in the large $\Delta$ limit. 
However, when $t_\text{v}/\Delta \ll 1$ we have $\xi \gtrsim L$ and saturation of IPR is found again for the same reason as Fig. \ref{fig-full-localized-chain} (b). Combining these two power-law dependence will yields $\xi^{-1} \propto (t_\text{v}/\Delta)^4$ in Eq. \ref{eq-loc-length}, using IPR$\propto 1/\xi$. 

To more accurately describe localization length as a function of eigenvalue $E$, 
we then derive an effective Hamiltonian as  
\begin{eqnarray}
H_{\text{eff}} = \sum_m t a_{m+1,0}^{\dagger}a_{m,0} + \mathrm{h.c.} + \sum_m (\Delta + W_m) a_{m,0}^{\dagger}a_{m,0},
\label{eq-couple-full-localized} 
\end{eqnarray}
where $W_m =  t_\text{v}^2/(\Delta - V_{m,1})$ for states with $E \sim \Delta$. A direct calculation shows that
\begin{eqnarray}
   \langle W_m \rangle &&= {1\over V}\int_{-V/2}^{V/2} W_m dV_{m,1}= \frac{t_\text{v}^2}{V}\ln(\frac{2|\Delta|/V + 1}{ 2|\Delta|/V -1}), \nonumber \\
   \langle W_m W_n \rangle &&= \frac{4t_\text{v}^4}{4\Delta^2 - V^2} \delta_{m,n},
\end{eqnarray} 
with $\langle \cdot \rangle$ represents its disorder averaged value. The variance of $W_m$ can be written as 
\begin{equation}
   v^2 = \langle W_m^2 \rangle - \langle W_m \rangle^2 = t_\text{v}^4(\frac{4}{4\Delta^2-V^2} 
 - f(\Delta, V)),
\end{equation}
with $f(\Delta, V) = (\frac{1}{V}\ln(\frac{2|\Delta|/V + 1}{ 2|\Delta|/V -1}))^2$.
It yields  
\begin{equation} 
v^2 = \frac{ t_\text{v}^4 V^2}{12\Delta^4} + \frac{11 t_\text{v}^4 V^4}{360 \Delta^6} + \mathcal{O}(\Delta^{-6}),
\end{equation}
when $\Delta \gg V$. The leading term yields the localization length in Eq. \ref{eq-loc-length} with a suppressed prefactor of  $t_\text{v}^4/\Delta^4$, which is numerically confirmed in Fig. \ref{fig-full-localized-chain} in a large system. This completes the proof of Eq. \ref{eq-loc-length}. 

\begin{figure}[htbp]
\centering
\includegraphics[width=0.35\textwidth]{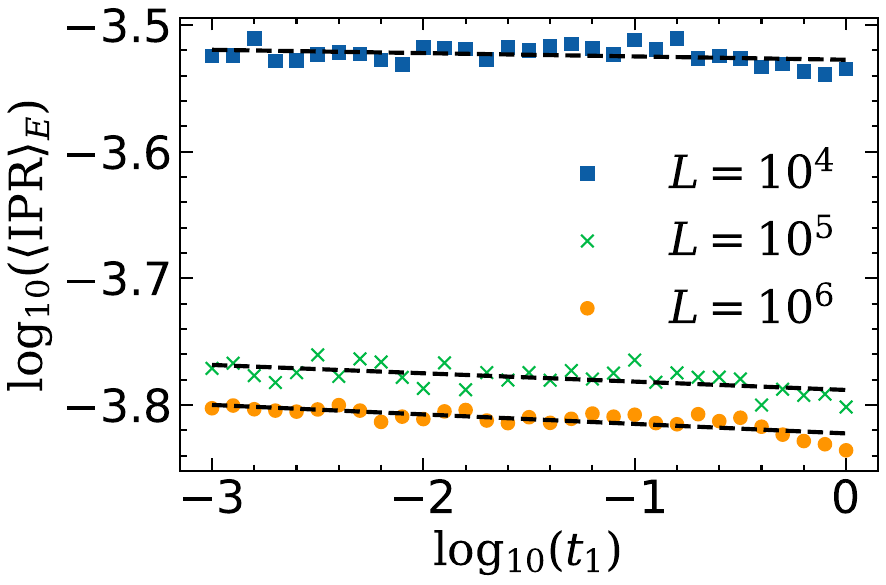}
\caption{
The logarithm of IPR versus the intra-chain hopping $t_1$, with $\Delta = -10$ and $t_\text{v} = 1$ for fixed energy $E = -10$. The black dashed lines are linear fitting with  $\log_{10}(\langle \text{IPR} \rangle_E) \sim \nu \log_{10}(t_1)$, with $\nu=-0.0026$ for $L=10^4$, $\nu = -0.0066$ for $L=10^5$, and $\nu = -0.0075$ for $L=10^5$ 
}
\label{fig-hopping}
\end{figure}

\begin{figure}[htbp]
\centering
\includegraphics[width=0.32\textwidth]{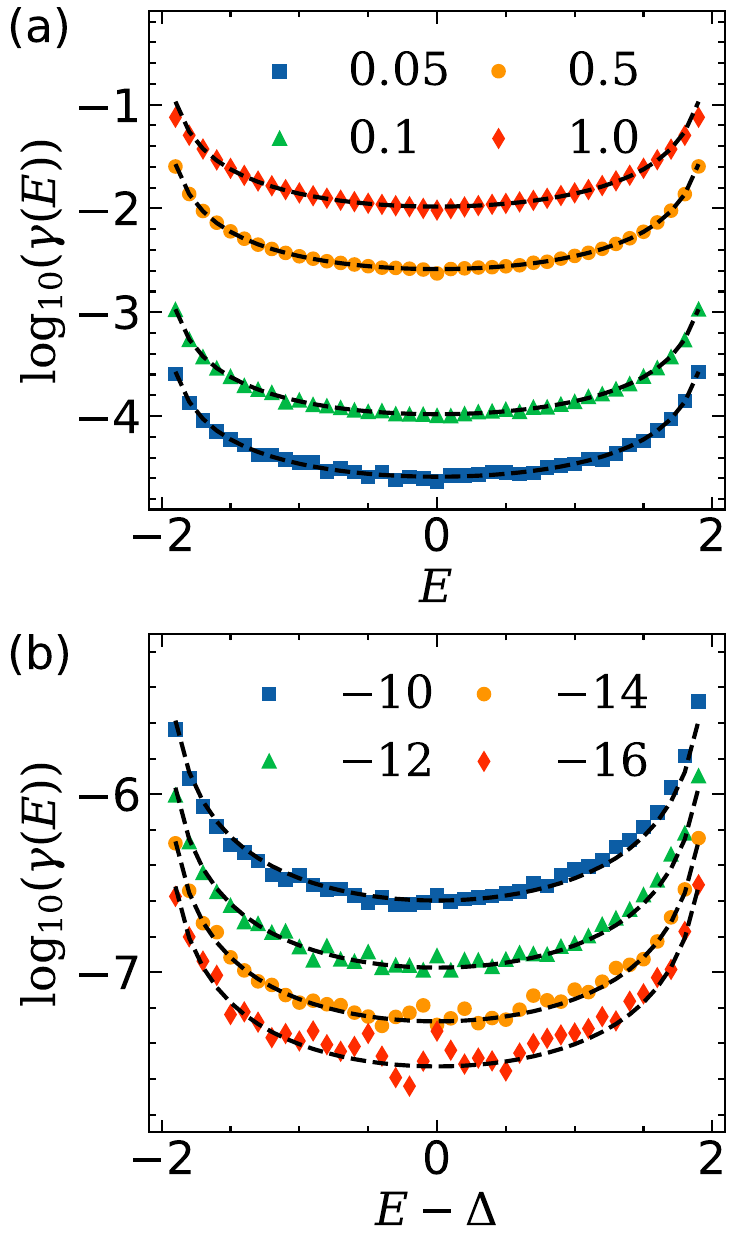}
\caption{
The logarithm of the Lyapunov exponent $\gamma(E)$ versus the energy $E$.
(a) Results in a single chain with different disorder strengths $V$, with
corresponding localization length by Eq. \ref{eq-thouless-expression}.
(b) Results in the un-overlapped regime in the coupled disordered model in 
Eq. \ref{eq-model} with different energy shifts $\Delta$ with $t_0 = t_1 = t$. The
localization length is given by Eq. \ref{eq-fullloclength}. In the transfer
matrix calculation we have used $L = 2\times10^9$.}
\label{fig-loc-length-compare}
\end{figure}

The previous conclusion is based on the minimal model with $t_1=0$.
We then move to examine the effect of hopping $t_1$ in the disordered chain, which is expected to extend the wave functions and hence suppressed the localization length in the free chain. 
Thus it is crucial to ask to what extent $t_1$ can influence the localization length $\xi$.
To this end, we fix $\Delta$ and $t_\text{v}$ and change the value of $t_1$, and the results of IPR against $t_1$ are presented in Fig. \ref{fig-hopping}.
We find that the IPR is slightly decreased with the increasing of $t_1$, indicating that $t_1$ is not the essential term for the localization of the free chain.
Therefore, we expect that Eq. \ref{eq-loc-length} serves as a good approximation for the localization length even with finite $t_1$, which accounts for the excellent agreement between the numerical and theoretical results in Fig. \ref{fig-localization-length} (a) and Fig. \ref{fig-loc-length-compare}. 
Finally, we present the relation between the localization length as a function of energy $E$ in a single disordered chain and in a coupled disordered model in Fig. \ref{fig-loc-length-compare}, which further confirms the empirical formulas of Eq. \ref{eq-thouless-expression} and Eq. \ref{eq-loc-length}. In Fig. \ref{fig-localization-length} (b), we have used 
\begin{equation}
\xi^{-1}(E) = (\frac{4}{4\Delta^2-V^2} 
 - f(\Delta, V))\frac{t_\text{v}^4}{8t^2 - 2(E-\Delta)^2},
\label{eq-fullloclength}
\end{equation}
which yields Eq. \ref{eq-loc-length} in the large $\Delta$ limit. We point out that for finite $\Delta$, the higher-order term $\mathcal{O}(\Delta^{-4})$ in $f(\Delta, V)$ is important. Therefore, while the localization length in the overlapped and un-overlapped spectra exhibits distinct behaviors with all wave functions being localized, it is not a phase transition process. 

\textit{Conclusion and discussion}: 
In this work, we present a coupled disordered model by coupling a disordered chain with a free chain, where the localization lengths in the overlapped and un-overlapped regimes differ by several orders of magnitude. 
In the overlapped regime, the states from the free chain are localized by resonant coupling between the localized and extended chains.
However, in the un-overlapped regime of the free chain, while the states are still localized by the general theorem, they exhibit suppressed localization with a prefactor of $ t_\text{v}^4 /\Delta^4$. 
We find that the inter-chain coupling, disorder strength, and energy shift play a leading role in localization, yet the effect of intra-chain hopping $t_1$ in the disordered chain is not significant.
The results presented in this work are in 1d disordered models, and extending this research to higher dimensional models \citep{Tarquini2017Critical, Jani2005Mean-field, Avishai2002New} and many-body models \citep{Nandkishore2015Many-body, Alet2018Many-body, Abanin2019Manybody} is also intriguing, in which we expect the overlapped regime and un-overlapped regime will also exhibit totally different behaviors \citep{Lin2023Singleparticle}.
Furthermore, for a system with a large localization length in the higher-dimensional models, it may call for much more advanced numerical methods. 

These results can be readily confirmed in the states-of-art experiments with ultracold atoms \citep{Roati2008Anderson,gross2017quantum, darkwah2022probing, Mandel2003Coherent, Yang2017Spin-dependent}, in which the two chains can be realized by the hyperfine states. The inter-chain coupling can be realized by Raman coupling and their potential shift is a natural consequence of detuning and Zeeman field. 
In these systems, the wave functions in each chain can be independently realized in the limit $t_v \sim 0$, and their localization can be measured individually using the time-of-flight imaging technique. In recent years, AL in disordered systems has been an important direction in ultracold atoms and huge progress has already been achieved \citep{Roati2008Anderson,  Skipetrov2008Anderson, Kohlert2019Observation, Alex2021Interaction, Wang2022Observation, white2020observation, Dikopoltsev2022Observation} and we expect the experimental confirmation of these results can provide perspicuous evidences for the dilemma of (I) and (II). 

Finally, it is necessary to emphasize that the disordered potential (with short-range correlation) has totally different features from the incommensurate potential.
In the coupled free chain and the incommensurate chain, without the guarantee of the general theorem, one can realize a critical phase in the overlapped spectra \cite{Lin2022General}, in which the overlapped and un-overlapped spectra also exhibit distinct behaviors in localization. 
The similar critical phase by coupling of extended and localized states in the Floquet model with incommensurate potential has also been presented by Roy {\it et al} in Ref. \citep{Roy2018Multifractality}. Here, we present a much-simplified model, which can be solved analytically in the limiting condition, in the hope that these intriguing results to be found in the more complicated coupled many-body models and coupled random matrices \cite{Lin2023Model}. 

\textit{Acknowledgments}: 
This work is supported by the National Natural Science Foundation of China (NSFC) with No. 11774328, and Innovation Program for Quantum Science and Technology (No. 2021ZD0301200 and No. 2021ZD0301500).

\bibliography{ref}

\begin{thebibliography}{64}%
\makeatletter
\providecommand \@ifxundefined [1]{%
 \@ifx{#1\undefined}
}%
\providecommand \@ifnum [1]{%
 \ifnum #1\expandafter \@firstoftwo
 \else \expandafter \@secondoftwo
 \fi
}%
\providecommand \@ifx [1]{%
 \ifx #1\expandafter \@firstoftwo
 \else \expandafter \@secondoftwo
 \fi
}%
\providecommand \natexlab [1]{#1}%
\providecommand \enquote  [1]{``#1''}%
\providecommand \bibnamefont  [1]{#1}%
\providecommand \bibfnamefont [1]{#1}%
\providecommand \citenamefont [1]{#1}%
\providecommand \href@noop [0]{\@secondoftwo}%
\providecommand \href [0]{\begingroup \@sanitize@url \@href}%
\providecommand \@href[1]{\@@startlink{#1}\@@href}%
\providecommand \@@href[1]{\endgroup#1\@@endlink}%
\providecommand \@sanitize@url [0]{\catcode `\\12\catcode `\$12\catcode
  `\&12\catcode `\#12\catcode `\^12\catcode `\_12\catcode `\%12\relax}%
\providecommand \@@startlink[1]{}%
\providecommand \@@endlink[0]{}%
\providecommand \url  [0]{\begingroup\@sanitize@url \@url }%
\providecommand \@url [1]{\endgroup\@href {#1}{\urlprefix }}%
\providecommand \urlprefix  [0]{URL }%
\providecommand \Eprint [0]{\href }%
\providecommand \doibase [0]{http://dx.doi.org/}%
\providecommand \selectlanguage [0]{\@gobble}%
\providecommand \bibinfo  [0]{\@secondoftwo}%
\providecommand \bibfield  [0]{\@secondoftwo}%
\providecommand \translation [1]{[#1]}%
\providecommand \BibitemOpen [0]{}%
\providecommand \bibitemStop [0]{}%
\providecommand \bibitemNoStop [0]{.\EOS\space}%
\providecommand \EOS [0]{\spacefactor3000\relax}%
\providecommand \BibitemShut  [1]{\csname bibitem#1\endcsname}%
\let\auto@bib@innerbib\@empty
\bibitem [{\citenamefont {Anderson}(1958)}]{anderson_absence_1958}%
  \BibitemOpen
  \bibfield  {author} {\bibinfo {author} {\bibfnamefont {P.~W.}\ \bibnamefont
  {Anderson}},\ }\bibfield  {title} {\enquote {\bibinfo {title} {Absence of
  diffusion in certain random lattices},}\ }\href {\doibase
  10.1103/PhysRev.109.1492} {\bibfield  {journal} {\bibinfo  {journal} {Phys.
  Rev.}\ }\textbf {\bibinfo {volume} {109}},\ \bibinfo {pages} {1492} (\bibinfo
  {year} {1958})}\BibitemShut {NoStop}%
\bibitem [{\citenamefont {Lee}\ and\ \citenamefont
  {Ramakrishnan}(1985)}]{Lee1985Disordered}%
  \BibitemOpen
  \bibfield  {author} {\bibinfo {author} {\bibfnamefont {Patrick}\ \bibnamefont
  {Lee}}\ and\ \bibinfo {author} {\bibfnamefont {T.~V.}\ \bibnamefont
  {Ramakrishnan}},\ }\bibfield  {title} {\enquote {\bibinfo {title} {Disordered
  electronic systems},}\ }\href {https://doi.org/10.1103/RevModPhys.57.287}
  {\bibfield  {journal} {\bibinfo  {journal} {Rev. Mod. Phys.}\ }\textbf
  {\bibinfo {volume} {57}},\ \bibinfo {pages} {287} (\bibinfo {year}
  {1985})}\BibitemShut {NoStop}%
\bibitem [{\citenamefont {Evers}\ and\ \citenamefont
  {Mirlin}(2008)}]{Evers2008Anderson}%
  \BibitemOpen
  \bibfield  {author} {\bibinfo {author} {\bibfnamefont {Ferdinand}\
  \bibnamefont {Evers}}\ and\ \bibinfo {author} {\bibfnamefont {Alexander~D.}\
  \bibnamefont {Mirlin}},\ }\bibfield  {title} {\enquote {\bibinfo {title}
  {Anderson transitions},}\ }\href {\doibase 10.1103/RevModPhys.80.1355}
  {\bibfield  {journal} {\bibinfo  {journal} {Rev. Mod. Phys.}\ }\textbf
  {\bibinfo {volume} {80}},\ \bibinfo {pages} {1355} (\bibinfo {year}
  {2008})}\BibitemShut {NoStop}%
\bibitem [{\citenamefont {Lagendijk}\ \emph {et~al.}(2009)\citenamefont
  {Lagendijk}, \citenamefont {Tiggelen},\ and\ \citenamefont
  {Wiersma}}]{Lagendijk2009Fifty}%
  \BibitemOpen
  \bibfield  {author} {\bibinfo {author} {\bibfnamefont {Ad}~\bibnamefont
  {Lagendijk}}, \bibinfo {author} {\bibfnamefont {Bart~van}\ \bibnamefont
  {Tiggelen}}, \ and\ \bibinfo {author} {\bibfnamefont {Diederik~S.}\
  \bibnamefont {Wiersma}},\ }\bibfield  {title} {\enquote {\bibinfo {title}
  {{Fifty years of Anderson localization}},}\ }\href {\doibase
  10.1063/1.3206091} {\bibfield  {journal} {\bibinfo  {journal} {Physics
  Today}\ }\textbf {\bibinfo {volume} {62}},\ \bibinfo {pages} {24} (\bibinfo
  {year} {2009})}\BibitemShut {NoStop}%
\bibitem [{\citenamefont {Roati}\ \emph {et~al.}(2008)\citenamefont {Roati},
  \citenamefont {D’Errico}, \citenamefont {Fallani}, \citenamefont {Fattori},
  \citenamefont {Fort}, \citenamefont {Zaccanti}, \citenamefont {Modugno},
  \citenamefont {Modugno},\ and\ \citenamefont {Inguscio}}]{Roati2008Anderson}%
  \BibitemOpen
  \bibfield  {author} {\bibinfo {author} {\bibfnamefont {Giacomo}\ \bibnamefont
  {Roati}}, \bibinfo {author} {\bibfnamefont {Chiara}\ \bibnamefont
  {D’Errico}}, \bibinfo {author} {\bibfnamefont {Leonardo}\ \bibnamefont
  {Fallani}}, \bibinfo {author} {\bibfnamefont {Marco}\ \bibnamefont
  {Fattori}}, \bibinfo {author} {\bibfnamefont {Chiara}\ \bibnamefont {Fort}},
  \bibinfo {author} {\bibfnamefont {Matteo}\ \bibnamefont {Zaccanti}}, \bibinfo
  {author} {\bibfnamefont {Giovanni}\ \bibnamefont {Modugno}}, \bibinfo
  {author} {\bibfnamefont {Michele}\ \bibnamefont {Modugno}}, \ and\ \bibinfo
  {author} {\bibfnamefont {Massimo}\ \bibnamefont {Inguscio}},\ }\bibfield
  {title} {\enquote {\bibinfo {title} {Anderson localization of a
  non-interacting {Bose}–{Einstein} condensate},}\ }\href {\doibase
  10.1038/nature07071} {\bibfield  {journal} {\bibinfo  {journal} {Nature}\
  }\textbf {\bibinfo {volume} {453}},\ \bibinfo {pages} {895} (\bibinfo {year}
  {2008})}\BibitemShut {NoStop}%
\bibitem [{\citenamefont {Segev}\ \emph {et~al.}(2013)\citenamefont {Segev},
  \citenamefont {Silberberg},\ and\ \citenamefont
  {Christodoulides}}]{Segev2013Anderson}%
  \BibitemOpen
  \bibfield  {author} {\bibinfo {author} {\bibfnamefont {Mordechai}\
  \bibnamefont {Segev}}, \bibinfo {author} {\bibfnamefont {Yaron}\ \bibnamefont
  {Silberberg}}, \ and\ \bibinfo {author} {\bibfnamefont {Demetrios~N.}\
  \bibnamefont {Christodoulides}},\ }\bibfield  {title} {\enquote {\bibinfo
  {title} {Anderson localization of light},}\ }\href {\doibase
  10.1038/nphoton.2013.30} {\bibfield  {journal} {\bibinfo  {journal} {Nat.
  Photonics}\ }\textbf {\bibinfo {volume} {7}},\ \bibinfo {pages} {197}
  (\bibinfo {year} {2013})}\BibitemShut {NoStop}%
\bibitem [{\citenamefont {Mott}\ and\ \citenamefont
  {Twose}(1961)}]{Mott1961Theory}%
  \BibitemOpen
  \bibfield  {author} {\bibinfo {author} {\bibfnamefont {N.F.}\ \bibnamefont
  {Mott}}\ and\ \bibinfo {author} {\bibfnamefont {W.D.}\ \bibnamefont
  {Twose}},\ }\bibfield  {title} {\enquote {\bibinfo {title} {The theory of
  impurity conduction},}\ }\href {\doibase 10.1080/00018736100101271}
  {\bibfield  {journal} {\bibinfo  {journal} {Adv. Phys.}\ }\textbf {\bibinfo
  {volume} {10}},\ \bibinfo {pages} {107} (\bibinfo {year} {1961})}\BibitemShut
  {NoStop}%
\bibitem [{\citenamefont {Abrahams}\ \emph {et~al.}(1979)\citenamefont
  {Abrahams}, \citenamefont {Anderson}, \citenamefont {Licciardello},\ and\
  \citenamefont {Ramakrishnan}}]{Abrahams1979Scaling}%
  \BibitemOpen
  \bibfield  {author} {\bibinfo {author} {\bibfnamefont {E.}~\bibnamefont
  {Abrahams}}, \bibinfo {author} {\bibfnamefont {P.~W.}\ \bibnamefont
  {Anderson}}, \bibinfo {author} {\bibfnamefont {D.~C.}\ \bibnamefont
  {Licciardello}}, \ and\ \bibinfo {author} {\bibfnamefont {T.~V.}\
  \bibnamefont {Ramakrishnan}},\ }\bibfield  {title} {\enquote {\bibinfo
  {title} {Scaling {Theory} of {Localization}: {Absence} of {Quantum}
  {Diffusion} in {Two} {Dimensions}},}\ }\href {\doibase
  10.1103/PhysRevLett.42.673} {\bibfield  {journal} {\bibinfo  {journal} {Phys.
  Rev. Lett.}\ }\textbf {\bibinfo {volume} {42}},\ \bibinfo {pages} {673}
  (\bibinfo {year} {1979})}\BibitemShut {NoStop}%
\bibitem [{\citenamefont {Ishii}(1973)}]{Ishii1973Localization}%
  \BibitemOpen
  \bibfield  {author} {\bibinfo {author} {\bibfnamefont {Kazushige}\
  \bibnamefont {Ishii}},\ }\bibfield  {title} {\enquote {\bibinfo {title}
  {Localization of {Eigenstates} and {Transport} {Phenomena} in the
  {One}-{Dimensional} {Disordered} {System}},}\ }\href {\doibase
  10.1143/PTPS.53.77} {\bibfield  {journal} {\bibinfo  {journal} {Prog. Theo.
  Phys. Supp.}\ }\textbf {\bibinfo {volume} {53}},\ \bibinfo {pages} {77}
  (\bibinfo {year} {1973})}\BibitemShut {NoStop}%
\bibitem [{\citenamefont {Thouless}(1972)}]{Thouless1972Relation}%
  \BibitemOpen
  \bibfield  {author} {\bibinfo {author} {\bibfnamefont {D~J}\ \bibnamefont
  {Thouless}},\ }\bibfield  {title} {\enquote {\bibinfo {title} {A relation
  between the density of states and range of localization for one dimensional
  random systems},}\ }\href {\doibase 10.1088/0022-3719/5/1/010} {\bibfield
  {journal} {\bibinfo  {journal} {J. Phys. C: Solid State Phys}\ }\textbf
  {\bibinfo {volume} {5}},\ \bibinfo {pages} {77} (\bibinfo {year}
  {1972})}\BibitemShut {NoStop}%
\bibitem [{\citenamefont {Thouless}(1973)}]{Thouless1973Localization}%
  \BibitemOpen
  \bibfield  {author} {\bibinfo {author} {\bibfnamefont {D~J}\ \bibnamefont
  {Thouless}},\ }\bibfield  {title} {\enquote {\bibinfo {title} {Localization
  distance and mean free path in one-dimensional disordered systems},}\ }\href
  {\doibase 10.1088/0022-3719/6/3/002} {\bibfield  {journal} {\bibinfo
  {journal} {J. Phys. C: Solid State Phys.}\ }\textbf {\bibinfo {volume} {6}},\
  \bibinfo {pages} {L49} (\bibinfo {year} {1973})}\BibitemShut {NoStop}%
\bibitem [{\citenamefont {Thouless}\ and\ \citenamefont
  {Kirkpatrick}(1981)}]{Thouless1981Conductivity}%
  \BibitemOpen
  \bibfield  {author} {\bibinfo {author} {\bibfnamefont {D.~J.}\ \bibnamefont
  {Thouless}}\ and\ \bibinfo {author} {\bibfnamefont {S.}~\bibnamefont
  {Kirkpatrick}},\ }\bibfield  {title} {\enquote {\bibinfo {title}
  {Conductivity of the disordered linear chain},}\ }\href {\doibase
  10.1088/0022-3719/14/3/007} {\bibfield  {journal} {\bibinfo  {journal} {J.
  Phys. C: Solid State Phys.}\ }\textbf {\bibinfo {volume} {14}},\ \bibinfo
  {pages} {235} (\bibinfo {year} {1981})}\BibitemShut {NoStop}%
\bibitem [{\citenamefont {Heinrichs}(2002)}]{Heinrichs2002Localization}%
  \BibitemOpen
  \bibfield  {author} {\bibinfo {author} {\bibfnamefont {J.}~\bibnamefont
  {Heinrichs}},\ }\bibfield  {title} {\enquote {\bibinfo {title} {Localization
  from conductance in few-channel disordered wires},}\ }\href {\doibase
  10.1103/PhysRevB.66.155434} {\bibfield  {journal} {\bibinfo  {journal} {Phys.
  Rev. B}\ }\textbf {\bibinfo {volume} {66}},\ \bibinfo {pages} {155434}
  (\bibinfo {year} {2002})}\BibitemShut {NoStop}%
\bibitem [{\citenamefont {Biddle}\ \emph {et~al.}(2009)\citenamefont {Biddle},
  \citenamefont {Wang}, \citenamefont {Priour},\ and\ \citenamefont
  {Das~Sarma}}]{Biddle2009Localization}%
  \BibitemOpen
  \bibfield  {author} {\bibinfo {author} {\bibfnamefont {J.}~\bibnamefont
  {Biddle}}, \bibinfo {author} {\bibfnamefont {B.}~\bibnamefont {Wang}},
  \bibinfo {author} {\bibfnamefont {D.~J.}\ \bibnamefont {Priour}}, \ and\
  \bibinfo {author} {\bibfnamefont {S.}~\bibnamefont {Das~Sarma}},\ }\bibfield
  {title} {\enquote {\bibinfo {title} {Localization in one-dimensional
  incommensurate lattices beyond the {Aubry}-{Andr}\'e model},}\ }\href
  {\doibase 10.1103/PhysRevA.80.021603} {\bibfield  {journal} {\bibinfo
  {journal} {Phys. Rev. A}\ }\textbf {\bibinfo {volume} {80}},\ \bibinfo
  {pages} {021603} (\bibinfo {year} {2009})}\BibitemShut {NoStop}%
\bibitem [{\citenamefont {Biddle}\ and\ \citenamefont
  {Das~Sarma}(2010)}]{Biddle2010Predicted}%
  \BibitemOpen
  \bibfield  {author} {\bibinfo {author} {\bibfnamefont {J.}~\bibnamefont
  {Biddle}}\ and\ \bibinfo {author} {\bibfnamefont {S.}~\bibnamefont
  {Das~Sarma}},\ }\bibfield  {title} {\enquote {\bibinfo {title} {Predicted
  {Mobility} {Edges} in {One}-{Dimensional} {Incommensurate} {Optical}
  {Lattices}: {An} {Exactly} {Solvable} {Model} of {Anderson}
  {Localization}},}\ }\href {\doibase 10.1103/PhysRevLett.104.070601}
  {\bibfield  {journal} {\bibinfo  {journal} {Phys. Rev. Lett.}\ }\textbf
  {\bibinfo {volume} {104}},\ \bibinfo {pages} {070601} (\bibinfo {year}
  {2010})}\BibitemShut {NoStop}%
\bibitem [{\citenamefont {Ganeshan}\ \emph {et~al.}(2015)\citenamefont
  {Ganeshan}, \citenamefont {Pixley},\ and\ \citenamefont
  {Das~Sarma}}]{Ganeshan2015Nearest}%
  \BibitemOpen
  \bibfield  {author} {\bibinfo {author} {\bibfnamefont {Sriram}\ \bibnamefont
  {Ganeshan}}, \bibinfo {author} {\bibfnamefont {J.~H.}\ \bibnamefont
  {Pixley}}, \ and\ \bibinfo {author} {\bibfnamefont {S.}~\bibnamefont
  {Das~Sarma}},\ }\bibfield  {title} {\enquote {\bibinfo {title} {Nearest
  {Neighbor} {Tight} {Binding} {Models} with an {Exact} {Mobility} {Edge} in
  {One} {Dimension}},}\ }\href {\doibase 10.1103/PhysRevLett.114.146601}
  {\bibfield  {journal} {\bibinfo  {journal} {Phys. Rev. Lett.}\ }\textbf
  {\bibinfo {volume} {114}},\ \bibinfo {pages} {146601} (\bibinfo {year}
  {2015})}\BibitemShut {NoStop}%
\bibitem [{\citenamefont {Wang}\ \emph {et~al.}(2020)\citenamefont {Wang},
  \citenamefont {Xia}, \citenamefont {Zhang}, \citenamefont {Yao},
  \citenamefont {Chen}, \citenamefont {You}, \citenamefont {Zhou},\ and\
  \citenamefont {Liu}}]{Wang2020one-dimensional}%
  \BibitemOpen
  \bibfield  {author} {\bibinfo {author} {\bibfnamefont {Yucheng}\ \bibnamefont
  {Wang}}, \bibinfo {author} {\bibfnamefont {Xu}~\bibnamefont {Xia}}, \bibinfo
  {author} {\bibfnamefont {Long}\ \bibnamefont {Zhang}}, \bibinfo {author}
  {\bibfnamefont {Hepeng}\ \bibnamefont {Yao}}, \bibinfo {author}
  {\bibfnamefont {Shu}\ \bibnamefont {Chen}}, \bibinfo {author} {\bibfnamefont
  {Jiangong}\ \bibnamefont {You}}, \bibinfo {author} {\bibfnamefont
  {Qi}~\bibnamefont {Zhou}}, \ and\ \bibinfo {author} {\bibfnamefont
  {Xiong-Jun}\ \bibnamefont {Liu}},\ }\bibfield  {title} {\enquote {\bibinfo
  {title} {One-{Dimensional} {Quasiperiodic} {Mosaic} {Lattice} with {Exact}
  {Mobility} {Edges}},}\ }\href {\doibase 10.1103/PhysRevLett.125.196604}
  {\bibfield  {journal} {\bibinfo  {journal} {Phys. Rev. Lett.}\ }\textbf
  {\bibinfo {volume} {125}},\ \bibinfo {pages} {196604} (\bibinfo {year}
  {2020})}\BibitemShut {NoStop}%
\bibitem [{\citenamefont {Izrailev}\ and\ \citenamefont
  {Krokhin}(1999)}]{Izrailev1999Localization}%
  \BibitemOpen
  \bibfield  {author} {\bibinfo {author} {\bibfnamefont {F.~M.}\ \bibnamefont
  {Izrailev}}\ and\ \bibinfo {author} {\bibfnamefont {A.~A.}\ \bibnamefont
  {Krokhin}},\ }\bibfield  {title} {\enquote {\bibinfo {title} {Localization
  and the mobility edge in one-dimensional potentials with correlated
  disorder},}\ }\href {\doibase 10.1103/PhysRevLett.82.4062} {\bibfield
  {journal} {\bibinfo  {journal} {Phys. Rev. Lett.}\ }\textbf {\bibinfo
  {volume} {82}},\ \bibinfo {pages} {4062} (\bibinfo {year}
  {1999})}\BibitemShut {NoStop}%
\bibitem [{\citenamefont {de~Moura}\ and\ \citenamefont
  {Lyra}(1998)}]{Moura1998Delocalization}%
  \BibitemOpen
  \bibfield  {author} {\bibinfo {author} {\bibfnamefont {Francisco A. B.~F.}\
  \bibnamefont {de~Moura}}\ and\ \bibinfo {author} {\bibfnamefont {Marcelo~L.}\
  \bibnamefont {Lyra}},\ }\bibfield  {title} {\enquote {\bibinfo {title}
  {Delocalization in the 1d anderson model with long-range correlated
  disorder},}\ }\href {\doibase 10.1103/PhysRevLett.81.3735} {\bibfield
  {journal} {\bibinfo  {journal} {Phys. Rev. Lett.}\ }\textbf {\bibinfo
  {volume} {81}},\ \bibinfo {pages} {3735} (\bibinfo {year}
  {1998})}\BibitemShut {NoStop}%
\bibitem [{\citenamefont {Dietz}\ \emph {et~al.}(2011)\citenamefont {Dietz},
  \citenamefont {Kuhl}, \citenamefont {St\"ockmann}, \citenamefont {Makarov},\
  and\ \citenamefont {Izrailev}}]{Dietz2011Microwave}%
  \BibitemOpen
  \bibfield  {author} {\bibinfo {author} {\bibfnamefont {O.}~\bibnamefont
  {Dietz}}, \bibinfo {author} {\bibfnamefont {U.}~\bibnamefont {Kuhl}},
  \bibinfo {author} {\bibfnamefont {H.-J.}\ \bibnamefont {St\"ockmann}},
  \bibinfo {author} {\bibfnamefont {N.~M.}\ \bibnamefont {Makarov}}, \ and\
  \bibinfo {author} {\bibfnamefont {F.~M.}\ \bibnamefont {Izrailev}},\
  }\bibfield  {title} {\enquote {\bibinfo {title} {Microwave realization of
  quasi-one-dimensional systems with correlated disorder},}\ }\href {\doibase
  10.1103/PhysRevB.83.134203} {\bibfield  {journal} {\bibinfo  {journal} {Phys.
  Rev. B}\ }\textbf {\bibinfo {volume} {83}},\ \bibinfo {pages} {134203}
  (\bibinfo {year} {2011})}\BibitemShut {NoStop}%
\bibitem [{\citenamefont {Delande}\ and\ \citenamefont
  {Orso}(2014)}]{Delande2014Mobility}%
  \BibitemOpen
  \bibfield  {author} {\bibinfo {author} {\bibfnamefont {Dominique}\
  \bibnamefont {Delande}}\ and\ \bibinfo {author} {\bibfnamefont {Giuliano}\
  \bibnamefont {Orso}},\ }\bibfield  {title} {\enquote {\bibinfo {title}
  {Mobility edge for cold atoms in laser speckle potentials},}\ }\href
  {\doibase 10.1103/PhysRevLett.113.060601} {\bibfield  {journal} {\bibinfo
  {journal} {Phys. Rev. Lett.}\ }\textbf {\bibinfo {volume} {113}},\ \bibinfo
  {pages} {060601} (\bibinfo {year} {2014})}\BibitemShut {NoStop}%
\bibitem [{\citenamefont {Johnston}\ and\ \citenamefont
  {Kramer}(1986)}]{Johnston1986Localization}%
  \BibitemOpen
  \bibfield  {author} {\bibinfo {author} {\bibfnamefont {R.}~\bibnamefont
  {Johnston}}\ and\ \bibinfo {author} {\bibfnamefont {B.}~\bibnamefont
  {Kramer}},\ }\bibfield  {title} {\enquote {\bibinfo {title} {Localization in
  one dimensional correlated random potentials},}\ }\href {\doibase
  10.1007/BF01303806} {\bibfield  {journal} {\bibinfo  {journal} {Z. Physik B -
  Condensed Matter}\ }\textbf {\bibinfo {volume} {63}},\ \bibinfo {pages} {273}
  (\bibinfo {year} {1986})}\BibitemShut {NoStop}%
\bibitem [{\citenamefont {Bordia}\ \emph {et~al.}(2016)\citenamefont {Bordia},
  \citenamefont {Lüschen}, \citenamefont {Hodgman}, \citenamefont {Schreiber},
  \citenamefont {Bloch},\ and\ \citenamefont {Schneider}}]{Bordia2016Coupling}%
  \BibitemOpen
  \bibfield  {author} {\bibinfo {author} {\bibfnamefont {Pranjal}\ \bibnamefont
  {Bordia}}, \bibinfo {author} {\bibfnamefont {Henrik~P.}\ \bibnamefont
  {Lüschen}}, \bibinfo {author} {\bibfnamefont {Sean~S.}\ \bibnamefont
  {Hodgman}}, \bibinfo {author} {\bibfnamefont {Michael}\ \bibnamefont
  {Schreiber}}, \bibinfo {author} {\bibfnamefont {Immanuel}\ \bibnamefont
  {Bloch}}, \ and\ \bibinfo {author} {\bibfnamefont {Ulrich}\ \bibnamefont
  {Schneider}},\ }\bibfield  {title} {\enquote {\bibinfo {title} {Coupling
  {Identical} one-dimensional {Many}-{Body} {Localized} {Systems}},}\ }\href
  {\doibase 10.1103/PhysRevLett.116.140401} {\bibfield  {journal} {\bibinfo
  {journal} {Phys. Rev. Lett.}\ }\textbf {\bibinfo {volume} {116}},\ \bibinfo
  {pages} {140401} (\bibinfo {year} {2016})}\BibitemShut {NoStop}%
\bibitem [{\citenamefont {Thiery}\ \emph {et~al.}(2018)\citenamefont {Thiery},
  \citenamefont {Huveneers}, \citenamefont {M{\"u}ller},\ and\ \citenamefont
  {De~Roeck}}]{Thiery2018Many}%
  \BibitemOpen
  \bibfield  {author} {\bibinfo {author} {\bibfnamefont {Thimoth{\'e}e}\
  \bibnamefont {Thiery}}, \bibinfo {author} {\bibfnamefont {Fran{\c{c}}ois}\
  \bibnamefont {Huveneers}}, \bibinfo {author} {\bibfnamefont {Markus}\
  \bibnamefont {M{\"u}ller}}, \ and\ \bibinfo {author} {\bibfnamefont
  {Wojciech}\ \bibnamefont {De~Roeck}},\ }\bibfield  {title} {\enquote
  {\bibinfo {title} {Many-body delocalization as a quantum avalanche},}\
  }\href@noop {} {\bibfield  {journal} {\bibinfo  {journal} {Phys. Rev. Lett.}\
  }\textbf {\bibinfo {volume} {121}},\ \bibinfo {pages} {140601} (\bibinfo
  {year} {2018})}\BibitemShut {NoStop}%
\bibitem [{\citenamefont {Chiew}\ \emph {et~al.}(2023)\citenamefont {Chiew},
  \citenamefont {Gong}, \citenamefont {Kwek},\ and\ \citenamefont
  {Lee}}]{Chiew2023Stability}%
  \BibitemOpen
  \bibfield  {author} {\bibinfo {author} {\bibfnamefont {Shao-Hen}\
  \bibnamefont {Chiew}}, \bibinfo {author} {\bibfnamefont {Jiangbin}\
  \bibnamefont {Gong}}, \bibinfo {author} {\bibfnamefont {Leong-Chuan}\
  \bibnamefont {Kwek}}, \ and\ \bibinfo {author} {\bibfnamefont {Chee-Kong}\
  \bibnamefont {Lee}},\ }\bibfield  {title} {\enquote {\bibinfo {title}
  {Stability and dynamics of many-body localized systems coupled to a small
  bath},}\ }\href {\doibase 10.1103/PhysRevB.107.224202} {\bibfield  {journal}
  {\bibinfo  {journal} {Phys. Rev. B}\ }\textbf {\bibinfo {volume} {107}},\
  \bibinfo {pages} {224202} (\bibinfo {year} {2023})}\BibitemShut {NoStop}%
\bibitem [{\citenamefont {L{\'e}onard}\ \emph {et~al.}(2023)\citenamefont
  {L{\'e}onard}, \citenamefont {Kim}, \citenamefont {Rispoli}, \citenamefont
  {Lukin}, \citenamefont {Schittko}, \citenamefont {Kwan}, \citenamefont
  {Demler}, \citenamefont {Sels},\ and\ \citenamefont
  {Greiner}}]{Leonard2023Probing}%
  \BibitemOpen
  \bibfield  {author} {\bibinfo {author} {\bibfnamefont {Julian}\ \bibnamefont
  {L{\'e}onard}}, \bibinfo {author} {\bibfnamefont {Sooshin}\ \bibnamefont
  {Kim}}, \bibinfo {author} {\bibfnamefont {Matthew}\ \bibnamefont {Rispoli}},
  \bibinfo {author} {\bibfnamefont {Alexander}\ \bibnamefont {Lukin}}, \bibinfo
  {author} {\bibfnamefont {Robert}\ \bibnamefont {Schittko}}, \bibinfo {author}
  {\bibfnamefont {Joyce}\ \bibnamefont {Kwan}}, \bibinfo {author}
  {\bibfnamefont {Eugene}\ \bibnamefont {Demler}}, \bibinfo {author}
  {\bibfnamefont {Dries}\ \bibnamefont {Sels}}, \ and\ \bibinfo {author}
  {\bibfnamefont {Markus}\ \bibnamefont {Greiner}},\ }\bibfield  {title}
  {\enquote {\bibinfo {title} {Probing the onset of quantum avalanches in a
  many-body localized system},}\ }\href
  {https://www.nature.com/articles/s41567-022-01887-3} {\bibfield  {journal}
  {\bibinfo  {journal} {Nat. Phys.}\ }\textbf {\bibinfo {volume} {19}},\
  \bibinfo {pages} {481} (\bibinfo {year} {2023})}\BibitemShut {NoStop}%
\bibitem [{\citenamefont {Ha}\ \emph {et~al.}(2023)\citenamefont {Ha},
  \citenamefont {Morningstar},\ and\ \citenamefont
  {Huse}}]{Hyunsoo2023Many-Body}%
  \BibitemOpen
  \bibfield  {author} {\bibinfo {author} {\bibfnamefont {Hyunsoo}\ \bibnamefont
  {Ha}}, \bibinfo {author} {\bibfnamefont {Alan}\ \bibnamefont {Morningstar}},
  \ and\ \bibinfo {author} {\bibfnamefont {David~A.}\ \bibnamefont {Huse}},\
  }\bibfield  {title} {\enquote {\bibinfo {title} {Many-body resonances in the
  avalanche instability of many-body localization},}\ }\href {\doibase
  10.1103/PhysRevLett.130.250405} {\bibfield  {journal} {\bibinfo  {journal}
  {Phys. Rev. Lett.}\ }\textbf {\bibinfo {volume} {130}},\ \bibinfo {pages}
  {250405} (\bibinfo {year} {2023})}\BibitemShut {NoStop}%
\bibitem [{\citenamefont {Brouwer}\ \emph {et~al.}(1998)\citenamefont
  {Brouwer}, \citenamefont {Mudry}, \citenamefont {Simons},\ and\ \citenamefont
  {Altland}}]{Brouwer1998Delocalization}%
  \BibitemOpen
  \bibfield  {author} {\bibinfo {author} {\bibfnamefont {P.~W.}\ \bibnamefont
  {Brouwer}}, \bibinfo {author} {\bibfnamefont {C.}~\bibnamefont {Mudry}},
  \bibinfo {author} {\bibfnamefont {B.~D.}\ \bibnamefont {Simons}}, \ and\
  \bibinfo {author} {\bibfnamefont {A.}~\bibnamefont {Altland}},\ }\bibfield
  {title} {\enquote {\bibinfo {title} {Delocalization in {Coupled}
  {One}-{Dimensional} {Chains}},}\ }\href {\doibase 10.1103/PhysRevLett.81.862}
  {\bibfield  {journal} {\bibinfo  {journal} {Phys. Rev. Lett.}\ }\textbf
  {\bibinfo {volume} {81}},\ \bibinfo {pages} {862} (\bibinfo {year}
  {1998})}\BibitemShut {NoStop}%
\bibitem [{\citenamefont {Mott}(1987)}]{Mott1987Mobility}%
  \BibitemOpen
  \bibfield  {author} {\bibinfo {author} {\bibfnamefont {N}~\bibnamefont
  {Mott}},\ }\bibfield  {title} {\enquote {\bibinfo {title} {The mobility edge
  since 1967},}\ }\href {\doibase 10.1088/0022-3719/20/21/008} {\bibfield
  {journal} {\bibinfo  {journal} {J. Phys. Condens. Matter}\ }\textbf {\bibinfo
  {volume} {20}},\ \bibinfo {pages} {3075} (\bibinfo {year}
  {1987})}\BibitemShut {NoStop}%
\bibitem [{\citenamefont {Beenakker}(1997)}]{Beenakker1997Random-matrix}%
  \BibitemOpen
  \bibfield  {author} {\bibinfo {author} {\bibfnamefont {C.~W.~J.}\
  \bibnamefont {Beenakker}},\ }\bibfield  {title} {\enquote {\bibinfo {title}
  {Random-matrix theory of quantum transport},}\ }\href {\doibase
  10.1103/RevModPhys.69.731} {\bibfield  {journal} {\bibinfo  {journal} {Rev.
  Mod. Phys.}\ }\textbf {\bibinfo {volume} {69}},\ \bibinfo {pages} {731}
  (\bibinfo {year} {1997})}\BibitemShut {NoStop}%
\bibitem [{\citenamefont {Belitz}\ and\ \citenamefont
  {Kirkpatrick}(1994)}]{Belitz1994Anderson}%
  \BibitemOpen
  \bibfield  {author} {\bibinfo {author} {\bibfnamefont {D.}~\bibnamefont
  {Belitz}}\ and\ \bibinfo {author} {\bibfnamefont {T.~R.}\ \bibnamefont
  {Kirkpatrick}},\ }\bibfield  {title} {\enquote {\bibinfo {title} {{The
  Anderson-Mott transition}},}\ }\href {\doibase 10.1103/RevModPhys.66.261}
  {\bibfield  {journal} {\bibinfo  {journal} {Rev. Mod. Phys.}\ }\textbf
  {\bibinfo {volume} {66}},\ \bibinfo {pages} {261--380} (\bibinfo {year}
  {1994})}\BibitemShut {NoStop}%
\bibitem [{\citenamefont {Thouless}(1974)}]{thouless1974electrons}%
  \BibitemOpen
  \bibfield  {author} {\bibinfo {author} {\bibfnamefont {D.~J.}\ \bibnamefont
  {Thouless}},\ }\bibfield  {title} {\enquote {\bibinfo {title} {Electrons in
  disordered systems and the theory of localization},}\ }\href {\doibase
  10.1016/0370-1573(74)90029-5} {\bibfield  {journal} {\bibinfo  {journal}
  {Phys. Rep.}\ }\textbf {\bibinfo {volume} {13}},\ \bibinfo {pages} {93}
  (\bibinfo {year} {1974})}\BibitemShut {NoStop}%
\bibitem [{\citenamefont {Efetov}(1983)}]{efetov1983supersymmetry}%
  \BibitemOpen
  \bibfield  {author} {\bibinfo {author} {\bibfnamefont {K.B.}\ \bibnamefont
  {Efetov}},\ }\bibfield  {title} {\enquote {\bibinfo {title} {Supersymmetry
  and theory of disordered metals},}\ }\href {\doibase
  10.1080/00018738300101531} {\bibfield  {journal} {\bibinfo  {journal}
  {Advances in Physics}\ }\textbf {\bibinfo {volume} {32}},\ \bibinfo {pages}
  {53} (\bibinfo {year} {1983})}\BibitemShut {NoStop}%
\bibitem [{\citenamefont {Theodorou}\ and\ \citenamefont
  {Cohen}(1976)}]{George1976Extended}%
  \BibitemOpen
  \bibfield  {author} {\bibinfo {author} {\bibfnamefont {George}\ \bibnamefont
  {Theodorou}}\ and\ \bibinfo {author} {\bibfnamefont {Morrel~H.}\ \bibnamefont
  {Cohen}},\ }\bibfield  {title} {\enquote {\bibinfo {title} {Extended states
  in a one-demensional system with off-diagonal disorder},}\ }\href {\doibase
  10.1103/PhysRevB.13.4597} {\bibfield  {journal} {\bibinfo  {journal} {Phys.
  Rev. B}\ }\textbf {\bibinfo {volume} {13}},\ \bibinfo {pages} {4597}
  (\bibinfo {year} {1976})}\BibitemShut {NoStop}%
\bibitem [{\citenamefont {Carmona}\ \emph {et~al.}(1987)\citenamefont
  {Carmona}, \citenamefont {Klein},\ and\ \citenamefont
  {Martinelli}}]{Carmona1987Anderson}%
  \BibitemOpen
  \bibfield  {author} {\bibinfo {author} {\bibfnamefont {Rene}\ \bibnamefont
  {Carmona}}, \bibinfo {author} {\bibfnamefont {Abel}\ \bibnamefont {Klein}}, \
  and\ \bibinfo {author} {\bibfnamefont {Fabio}\ \bibnamefont {Martinelli}},\
  }\bibfield  {title} {\enquote {\bibinfo {title} {Anderson localization for
  {Bernoulli} and other singular potentials},}\ }\href {\doibase
  10.1007/BF01210702} {\bibfield  {journal} {\bibinfo  {journal} {Commun.Math.
  Phys.}\ }\textbf {\bibinfo {volume} {108}},\ \bibinfo {pages} {41} (\bibinfo
  {year} {1987})}\BibitemShut {NoStop}%
\bibitem [{\citenamefont {Kunz}\ and\ \citenamefont
  {Souillard}(1980)}]{Kunz1980Sur}%
  \BibitemOpen
  \bibfield  {author} {\bibinfo {author} {\bibfnamefont {Hervé}\ \bibnamefont
  {Kunz}}\ and\ \bibinfo {author} {\bibfnamefont {Bernard}\ \bibnamefont
  {Souillard}},\ }\bibfield  {title} {\enquote {\bibinfo {title} {Sur le
  spectre des opérateurs aux différences finies aléatoires},}\ }\href
  {\doibase 10.1007/BF01942371} {\bibfield  {journal} {\bibinfo  {journal}
  {Commun.Math. Phys.}\ }\textbf {\bibinfo {volume} {78}},\ \bibinfo {pages}
  {201} (\bibinfo {year} {1980})}\BibitemShut {NoStop}%
\bibitem [{\citenamefont {Fröhlich}\ and\ \citenamefont
  {Spencer}(1983)}]{Frohlich1983Absence}%
  \BibitemOpen
  \bibfield  {author} {\bibinfo {author} {\bibfnamefont {Jürg}\ \bibnamefont
  {Fröhlich}}\ and\ \bibinfo {author} {\bibfnamefont {Thomas}\ \bibnamefont
  {Spencer}},\ }\bibfield  {title} {\enquote {\bibinfo {title} {Absence of
  diffusion in the {Anderson} tight binding model for large disorder or low
  energy},}\ }\href {\doibase 10.1007/BF01209475} {\bibfield  {journal}
  {\bibinfo  {journal} {Commun. Math. Phys.}\ }\textbf {\bibinfo {volume}
  {88}},\ \bibinfo {pages} {151} (\bibinfo {year} {1983})}\BibitemShut
  {NoStop}%
\bibitem [{\citenamefont {Pastur}(1980)}]{Pastur1980Spectral}%
  \BibitemOpen
  \bibfield  {author} {\bibinfo {author} {\bibfnamefont {L.~A.}\ \bibnamefont
  {Pastur}},\ }\bibfield  {title} {\enquote {\bibinfo {title} {Spectral
  properties of disordered systems in the one-body approximation},}\ }\href
  {https://link.springer.com/article/10.1007/BF01222516} {\bibfield  {journal}
  {\bibinfo  {journal} {Communications in Mathematical Physics}\ }\textbf
  {\bibinfo {volume} {75}},\ \bibinfo {pages} {179 -- 196} (\bibinfo {year}
  {1980})}\BibitemShut {NoStop}%
\bibitem [{Hof()}]{Hoffmanbook}%
  \BibitemOpen
  \href@noop {} {}\bibinfo {note} {Karl Heinz Hoffmann, Michael Schreiber
  (Eds.), Computational Physics, Springer Berlin Heidelberg, 1996.}\BibitemShut
  {Stop}%
\bibitem [{\citenamefont {Oseledets}(1968)}]{Oseledets1968Multiplicative}%
  \BibitemOpen
  \bibfield  {author} {\bibinfo {author} {\bibfnamefont {Valery~Iustinovich}\
  \bibnamefont {Oseledets}},\ }\bibfield  {title} {\enquote {\bibinfo {title}
  {A multiplicative ergodic theorem. characteristic lyapunov, exponents of
  dynamical systems},}\ }\href@noop {} {\bibfield  {journal} {\bibinfo
  {journal} {Trudy Moskovskogo Matematicheskogo Obshchestva}\ }\textbf
  {\bibinfo {volume} {19}},\ \bibinfo {pages} {179} (\bibinfo {year}
  {1968})}\BibitemShut {NoStop}%
\bibitem [{\citenamefont {Raghunathan}(1979)}]{Raghunathan1979Proof}%
  \BibitemOpen
  \bibfield  {author} {\bibinfo {author} {\bibfnamefont {Madabusi~S}\
  \bibnamefont {Raghunathan}},\ }\bibfield  {title} {\enquote {\bibinfo {title}
  {A proof of oseledec’s multiplicative ergodic theorem},}\ }\href@noop {}
  {\bibfield  {journal} {\bibinfo  {journal} {Israel J. Math.}\ }\textbf
  {\bibinfo {volume} {32}},\ \bibinfo {pages} {356} (\bibinfo {year}
  {1979})}\BibitemShut {NoStop}%
\bibitem [{\citenamefont {Pietracaprina}\ \emph {et~al.}(2018)\citenamefont
  {Pietracaprina}, \citenamefont {Mac{\'e}}, \citenamefont {Luitz},\ and\
  \citenamefont {Alet}}]{Pietracaprina2018Shift}%
  \BibitemOpen
  \bibfield  {author} {\bibinfo {author} {\bibfnamefont {Francesca}\
  \bibnamefont {Pietracaprina}}, \bibinfo {author} {\bibfnamefont {Nicolas}\
  \bibnamefont {Mac{\'e}}}, \bibinfo {author} {\bibfnamefont {David~J}\
  \bibnamefont {Luitz}}, \ and\ \bibinfo {author} {\bibfnamefont {Fabien}\
  \bibnamefont {Alet}},\ }\bibfield  {title} {\enquote {\bibinfo {title}
  {Shift-invert diagonalization of large many-body localizing spin chains},}\
  }\href {https://scipost.org/SciPostPhys.5.5.045} {\bibfield  {journal}
  {\bibinfo  {journal} {SciPost Phys.}\ }\textbf {\bibinfo {volume} {5}},\
  \bibinfo {pages} {045} (\bibinfo {year} {2018})}\BibitemShut {NoStop}%
\bibitem [{\citenamefont {Luitz}\ \emph {et~al.}(2015)\citenamefont {Luitz},
  \citenamefont {Laflorencie},\ and\ \citenamefont
  {Alet}}]{Luitz2015Many-body}%
  \BibitemOpen
  \bibfield  {author} {\bibinfo {author} {\bibfnamefont {David~J.}\
  \bibnamefont {Luitz}}, \bibinfo {author} {\bibfnamefont {Nicolas}\
  \bibnamefont {Laflorencie}}, \ and\ \bibinfo {author} {\bibfnamefont
  {Fabien}\ \bibnamefont {Alet}},\ }\bibfield  {title} {\enquote {\bibinfo
  {title} {Many-body localization edge in the random-field {Heisenberg}
  chain},}\ }\href {\doibase 10.1103/PhysRevB.91.081103} {\bibfield  {journal}
  {\bibinfo  {journal} {Phys. Rev. B}\ }\textbf {\bibinfo {volume} {91}},\
  \bibinfo {pages} {081103} (\bibinfo {year} {2015})}\BibitemShut {NoStop}%
\bibitem [{\citenamefont {Hiramoto}\ and\ \citenamefont
  {Kohmoto}(1989)}]{Hiramoto1989Scaling}%
  \BibitemOpen
  \bibfield  {author} {\bibinfo {author} {\bibfnamefont {Hisashi}\ \bibnamefont
  {Hiramoto}}\ and\ \bibinfo {author} {\bibfnamefont {Mahito}\ \bibnamefont
  {Kohmoto}},\ }\bibfield  {title} {\enquote {\bibinfo {title} {Scaling
  analysis of quasiperiodic systems: {Generalized} {Harper} model},}\ }\href
  {\doibase 10.1103/PhysRevB.40.8225} {\bibfield  {journal} {\bibinfo
  {journal} {Phys. Rev. B}\ }\textbf {\bibinfo {volume} {40}},\ \bibinfo
  {pages} {8225} (\bibinfo {year} {1989})}\BibitemShut {NoStop}%
\bibitem [{\citenamefont {Lin}\ \emph {et~al.}(2022)\citenamefont {Lin},
  \citenamefont {Chen}, \citenamefont {Guo},\ and\ \citenamefont
  {Gong}}]{Lin2022General}%
  \BibitemOpen
  \bibfield  {author} {\bibinfo {author} {\bibfnamefont {Xiaoshui}\
  \bibnamefont {Lin}}, \bibinfo {author} {\bibfnamefont {Xiaoman}\ \bibnamefont
  {Chen}}, \bibinfo {author} {\bibfnamefont {Guang-Can}\ \bibnamefont {Guo}}, \
  and\ \bibinfo {author} {\bibfnamefont {Ming}\ \bibnamefont {Gong}},\ }\href
  {\doibase 10.48550/ARXIV.2209.03060} {\enquote {\bibinfo {title} {The general
  approach to the critical phase with coupled quasiperiodic chains},}\ }
  (\bibinfo {year} {2022}),\ \bibinfo {note} {arXiv:2209.03060}\BibitemShut
  {NoStop}%
\bibitem [{\citenamefont {Tarquini}\ \emph {et~al.}(2017)\citenamefont
  {Tarquini}, \citenamefont {Biroli},\ and\ \citenamefont
  {Tarzia}}]{Tarquini2017Critical}%
  \BibitemOpen
  \bibfield  {author} {\bibinfo {author} {\bibfnamefont {E.}~\bibnamefont
  {Tarquini}}, \bibinfo {author} {\bibfnamefont {G.}~\bibnamefont {Biroli}}, \
  and\ \bibinfo {author} {\bibfnamefont {M.}~\bibnamefont {Tarzia}},\
  }\bibfield  {title} {\enquote {\bibinfo {title} {Critical properties of the
  {Anderson} localization transition and the high-dimensional limit},}\ }\href
  {\doibase 10.1103/PhysRevB.95.094204} {\bibfield  {journal} {\bibinfo
  {journal} {Phys. Rev. B}\ }\textbf {\bibinfo {volume} {95}},\ \bibinfo
  {pages} {094204} (\bibinfo {year} {2017})}\BibitemShut {NoStop}%
\bibitem [{\citenamefont {Jani\ifmmode~\check{s}\else \v{s}\fi{}}\ and\
  \citenamefont {Koloren\ifmmode~\check{c}\else
  \v{c}\fi{}}(2005)}]{Jani2005Mean-field}%
  \BibitemOpen
  \bibfield  {author} {\bibinfo {author} {\bibfnamefont {V.}~\bibnamefont
  {Jani\ifmmode~\check{s}\else \v{s}\fi{}}}\ and\ \bibinfo {author}
  {\bibfnamefont {J.}~\bibnamefont {Koloren\ifmmode~\check{c}\else
  \v{c}\fi{}}},\ }\bibfield  {title} {\enquote {\bibinfo {title} {{Mean-field
  theory of Anderson localization: Asymptotic solution in high spatial
  dimensions}},}\ }\href {\doibase 10.1103/PhysRevB.71.033103} {\bibfield
  {journal} {\bibinfo  {journal} {Phys. Rev. B}\ }\textbf {\bibinfo {volume}
  {71}},\ \bibinfo {pages} {033103} (\bibinfo {year} {2005})}\BibitemShut
  {NoStop}%
\bibitem [{\citenamefont {Avishai}\ and\ \citenamefont
  {Meir}(2002)}]{Avishai2002New}%
  \BibitemOpen
  \bibfield  {author} {\bibinfo {author} {\bibfnamefont {Yshai}\ \bibnamefont
  {Avishai}}\ and\ \bibinfo {author} {\bibfnamefont {Yigal}\ \bibnamefont
  {Meir}},\ }\bibfield  {title} {\enquote {\bibinfo {title} {New
  spin-orbit-induced universality class in the integer quantum hall regime},}\
  }\href {\doibase 10.1103/PhysRevLett.89.076602} {\bibfield  {journal}
  {\bibinfo  {journal} {Phys. Rev. Lett.}\ }\textbf {\bibinfo {volume} {89}},\
  \bibinfo {pages} {076602} (\bibinfo {year} {2002})}\BibitemShut {NoStop}%
\bibitem [{\citenamefont {Nandkishore}\ and\ \citenamefont
  {Huse}(2015)}]{Nandkishore2015Many-body}%
  \BibitemOpen
  \bibfield  {author} {\bibinfo {author} {\bibfnamefont {Rahul}\ \bibnamefont
  {Nandkishore}}\ and\ \bibinfo {author} {\bibfnamefont {David~A.}\
  \bibnamefont {Huse}},\ }\bibfield  {title} {\enquote {\bibinfo {title}
  {Many-{Body} {Localization} and {Thermalization} in {Quantum} {Statistical}
  {Mechanics}},}\ }\href {\doibase 10.1146/annurev-conmatphys-031214-014726}
  {\bibfield  {journal} {\bibinfo  {journal} {Annu. Rev. Condens. Matter
  Phys.}\ }\textbf {\bibinfo {volume} {6}},\ \bibinfo {pages} {15} (\bibinfo
  {year} {2015})}\BibitemShut {NoStop}%
\bibitem [{\citenamefont {Alet}\ and\ \citenamefont
  {Laflorencie}(2018)}]{Alet2018Many-body}%
  \BibitemOpen
  \bibfield  {author} {\bibinfo {author} {\bibfnamefont {Fabien}\ \bibnamefont
  {Alet}}\ and\ \bibinfo {author} {\bibfnamefont {Nicolas}\ \bibnamefont
  {Laflorencie}},\ }\bibfield  {title} {\enquote {\bibinfo {title} {Many-body
  localization: an introduction and selected topics},}\ }\href {\doibase
  10.1016/j.crhy.2018.03.003} {\bibfield  {journal} {\bibinfo  {journal}
  {Comptes Rendus Physique}\ }\textbf {\bibinfo {volume} {19}},\ \bibinfo
  {pages} {498} (\bibinfo {year} {2018})}\BibitemShut {NoStop}%
\bibitem [{\citenamefont {Abanin}\ \emph {et~al.}(2019)\citenamefont {Abanin},
  \citenamefont {Altman}, \citenamefont {Bloch},\ and\ \citenamefont
  {Serbyn}}]{Abanin2019Manybody}%
  \BibitemOpen
  \bibfield  {author} {\bibinfo {author} {\bibfnamefont {Dmitry~A.}\
  \bibnamefont {Abanin}}, \bibinfo {author} {\bibfnamefont {Ehud}\ \bibnamefont
  {Altman}}, \bibinfo {author} {\bibfnamefont {Immanuel}\ \bibnamefont
  {Bloch}}, \ and\ \bibinfo {author} {\bibfnamefont {Maksym}\ \bibnamefont
  {Serbyn}},\ }\bibfield  {title} {\enquote {\bibinfo {title} {Colloquium:
  Many-body localization, thermalization, and entanglement},}\ }\href {\doibase
  10.1103/RevModPhys.91.021001} {\bibfield  {journal} {\bibinfo  {journal}
  {Rev. Mod. Phys.}\ }\textbf {\bibinfo {volume} {91}},\ \bibinfo {pages}
  {021001} (\bibinfo {year} {2019})}\BibitemShut {NoStop}%
\bibitem [{\citenamefont {Lin}\ \emph {et~al.}(2023)\citenamefont {Lin},
  \citenamefont {Gong},\ and\ \citenamefont {Guo}}]{Lin2023Singleparticle}%
  \BibitemOpen
  \bibfield  {author} {\bibinfo {author} {\bibfnamefont {Xiaoshui}\
  \bibnamefont {Lin}}, \bibinfo {author} {\bibfnamefont {Ming}\ \bibnamefont
  {Gong}}, \ and\ \bibinfo {author} {\bibfnamefont {Guang-Can}\ \bibnamefont
  {Guo}},\ }\href@noop {} {\enquote {\bibinfo {title} {From single-particle to
  many-body mobility edges and the fate of overlapped spectra in coupled
  disorder models},}\ } (\bibinfo {year} {2023}),\ \Eprint
  {http://arxiv.org/abs/2307.01638} {arXiv:2307.01638} \BibitemShut {NoStop}%
\bibitem [{\citenamefont {Gross}\ and\ \citenamefont
  {Bloch}(2017)}]{gross2017quantum}%
  \BibitemOpen
  \bibfield  {author} {\bibinfo {author} {\bibfnamefont {Christian}\
  \bibnamefont {Gross}}\ and\ \bibinfo {author} {\bibfnamefont {Immanuel}\
  \bibnamefont {Bloch}},\ }\bibfield  {title} {\enquote {\bibinfo {title}
  {Quantum simulations with ultracold atoms in optical lattices},}\ }\href
  {\doibase 10.1126/science.aal3837} {\bibfield  {journal} {\bibinfo  {journal}
  {Science}\ }\textbf {\bibinfo {volume} {357}},\ \bibinfo {pages} {995}
  (\bibinfo {year} {2017})}\BibitemShut {NoStop}%
\bibitem [{\citenamefont {Darkwah~Oppong}\ \emph {et~al.}(2022)\citenamefont
  {Darkwah~Oppong}, \citenamefont {Pasqualetti}, \citenamefont {Bettermann},
  \citenamefont {Zechmann}, \citenamefont {Knap}, \citenamefont {Bloch},\ and\
  \citenamefont {F\"olling}}]{darkwah2022probing}%
  \BibitemOpen
  \bibfield  {author} {\bibinfo {author} {\bibfnamefont {N.}~\bibnamefont
  {Darkwah~Oppong}}, \bibinfo {author} {\bibfnamefont {G.}~\bibnamefont
  {Pasqualetti}}, \bibinfo {author} {\bibfnamefont {O.}~\bibnamefont
  {Bettermann}}, \bibinfo {author} {\bibfnamefont {P.}~\bibnamefont
  {Zechmann}}, \bibinfo {author} {\bibfnamefont {M.}~\bibnamefont {Knap}},
  \bibinfo {author} {\bibfnamefont {I.}~\bibnamefont {Bloch}}, \ and\ \bibinfo
  {author} {\bibfnamefont {S.}~\bibnamefont {F\"olling}},\ }\bibfield  {title}
  {\enquote {\bibinfo {title} {Probing transport and slow relaxation in the
  mass-imbalanced fermi-hubbard model},}\ }\href {\doibase
  10.1103/PhysRevX.12.031026} {\bibfield  {journal} {\bibinfo  {journal} {Phys.
  Rev. X}\ }\textbf {\bibinfo {volume} {12}},\ \bibinfo {pages} {031026}
  (\bibinfo {year} {2022})}\BibitemShut {NoStop}%
\bibitem [{\citenamefont {Mandel}\ \emph {et~al.}(2003)\citenamefont {Mandel},
  \citenamefont {Greiner}, \citenamefont {Widera}, \citenamefont {Rom},
  \citenamefont {Hänsch},\ and\ \citenamefont {Bloch}}]{Mandel2003Coherent}%
  \BibitemOpen
  \bibfield  {author} {\bibinfo {author} {\bibfnamefont {Olaf}\ \bibnamefont
  {Mandel}}, \bibinfo {author} {\bibfnamefont {Markus}\ \bibnamefont
  {Greiner}}, \bibinfo {author} {\bibfnamefont {Artur}\ \bibnamefont {Widera}},
  \bibinfo {author} {\bibfnamefont {Tim}\ \bibnamefont {Rom}}, \bibinfo
  {author} {\bibfnamefont {Theodor~W.}\ \bibnamefont {Hänsch}}, \ and\
  \bibinfo {author} {\bibfnamefont {Immanuel}\ \bibnamefont {Bloch}},\
  }\bibfield  {title} {\enquote {\bibinfo {title} {Coherent {Transport} of
  {Neutral} {Atoms} in {Spin}-{Dependent} {Optical} {Lattice} {Potentials}},}\
  }\href {\doibase 10.1103/PhysRevLett.91.010407} {\bibfield  {journal}
  {\bibinfo  {journal} {Phys. Rev. Lett.}\ }\textbf {\bibinfo {volume} {91}},\
  \bibinfo {pages} {010407} (\bibinfo {year} {2003})}\BibitemShut {NoStop}%
\bibitem [{\citenamefont {Yang}\ \emph {et~al.}(2017)\citenamefont {Yang},
  \citenamefont {Dai}, \citenamefont {Sun}, \citenamefont {Reingruber},
  \citenamefont {Yuan},\ and\ \citenamefont {Pan}}]{Yang2017Spin-dependent}%
  \BibitemOpen
  \bibfield  {author} {\bibinfo {author} {\bibfnamefont {Bing}\ \bibnamefont
  {Yang}}, \bibinfo {author} {\bibfnamefont {Han-Ning}\ \bibnamefont {Dai}},
  \bibinfo {author} {\bibfnamefont {Hui}\ \bibnamefont {Sun}}, \bibinfo
  {author} {\bibfnamefont {Andreas}\ \bibnamefont {Reingruber}}, \bibinfo
  {author} {\bibfnamefont {Zhen-Sheng}\ \bibnamefont {Yuan}}, \ and\ \bibinfo
  {author} {\bibfnamefont {Jian-Wei}\ \bibnamefont {Pan}},\ }\bibfield  {title}
  {\enquote {\bibinfo {title} {Spin-dependent optical superlattice},}\ }\href
  {\doibase 10.1103/PhysRevA.96.011602} {\bibfield  {journal} {\bibinfo
  {journal} {Phys. Rev. A}\ }\textbf {\bibinfo {volume} {96}},\ \bibinfo
  {pages} {011602} (\bibinfo {year} {2017})}\BibitemShut {NoStop}%
\bibitem [{\citenamefont {Skipetrov}\ \emph {et~al.}(2008)\citenamefont
  {Skipetrov}, \citenamefont {Minguzzi}, \citenamefont {van Tiggelen},\ and\
  \citenamefont {Shapiro}}]{Skipetrov2008Anderson}%
  \BibitemOpen
  \bibfield  {author} {\bibinfo {author} {\bibfnamefont {S.~E.}\ \bibnamefont
  {Skipetrov}}, \bibinfo {author} {\bibfnamefont {A.}~\bibnamefont {Minguzzi}},
  \bibinfo {author} {\bibfnamefont {B.~A.}\ \bibnamefont {van Tiggelen}}, \
  and\ \bibinfo {author} {\bibfnamefont {B.}~\bibnamefont {Shapiro}},\
  }\bibfield  {title} {\enquote {\bibinfo {title} {{Anderson Localization of a
  Bose-Einstein Condensate in a 3D Random Potential}},}\ }\href {\doibase
  10.1103/PhysRevLett.100.165301} {\bibfield  {journal} {\bibinfo  {journal}
  {Phys. Rev. Lett.}\ }\textbf {\bibinfo {volume} {100}},\ \bibinfo {pages}
  {165301} (\bibinfo {year} {2008})}\BibitemShut {NoStop}%
\bibitem [{\citenamefont {Kohlert}\ \emph {et~al.}(2019)\citenamefont
  {Kohlert}, \citenamefont {Scherg}, \citenamefont {Li}, \citenamefont
  {Lüschen}, \citenamefont {Das~Sarma}, \citenamefont {Bloch},\ and\
  \citenamefont {Aidelsburger}}]{Kohlert2019Observation}%
  \BibitemOpen
  \bibfield  {author} {\bibinfo {author} {\bibfnamefont {Thomas}\ \bibnamefont
  {Kohlert}}, \bibinfo {author} {\bibfnamefont {Sebastian}\ \bibnamefont
  {Scherg}}, \bibinfo {author} {\bibfnamefont {Xiao}\ \bibnamefont {Li}},
  \bibinfo {author} {\bibfnamefont {Henrik~P.}\ \bibnamefont {Lüschen}},
  \bibinfo {author} {\bibfnamefont {Sankar}\ \bibnamefont {Das~Sarma}},
  \bibinfo {author} {\bibfnamefont {Immanuel}\ \bibnamefont {Bloch}}, \ and\
  \bibinfo {author} {\bibfnamefont {Monika}\ \bibnamefont {Aidelsburger}},\
  }\bibfield  {title} {\enquote {\bibinfo {title} {Observation of {Many}-{Body}
  {Localization} in a {One}-{Dimensional} {System} with a {Single}-{Particle}
  {Mobility} {Edge}},}\ }\href {\doibase 10.1103/PhysRevLett.122.170403}
  {\bibfield  {journal} {\bibinfo  {journal} {Phys. Rev. Lett.}\ }\textbf
  {\bibinfo {volume} {122}},\ \bibinfo {pages} {170403} (\bibinfo {year}
  {2019})}\BibitemShut {NoStop}%
\bibitem [{\citenamefont {An}\ \emph {et~al.}(2021)\citenamefont {An},
  \citenamefont {Padavi\ifmmode~\acute{c}\else \'{c}\fi{}}, \citenamefont
  {Meier}, \citenamefont {Hegde}, \citenamefont {Ganeshan}, \citenamefont
  {Pixley}, \citenamefont {Vishveshwara},\ and\ \citenamefont
  {Gadway}}]{Alex2021Interaction}%
  \BibitemOpen
  \bibfield  {author} {\bibinfo {author} {\bibfnamefont {Fangzhao~Alex}\
  \bibnamefont {An}}, \bibinfo {author} {\bibfnamefont {Karmela}\ \bibnamefont
  {Padavi\ifmmode~\acute{c}\else \'{c}\fi{}}}, \bibinfo {author} {\bibfnamefont
  {Eric~J.}\ \bibnamefont {Meier}}, \bibinfo {author} {\bibfnamefont {Suraj}\
  \bibnamefont {Hegde}}, \bibinfo {author} {\bibfnamefont {Sriram}\
  \bibnamefont {Ganeshan}}, \bibinfo {author} {\bibfnamefont {J.~H.}\
  \bibnamefont {Pixley}}, \bibinfo {author} {\bibfnamefont {Smitha}\
  \bibnamefont {Vishveshwara}}, \ and\ \bibinfo {author} {\bibfnamefont
  {Bryce}\ \bibnamefont {Gadway}},\ }\bibfield  {title} {\enquote {\bibinfo
  {title} {{Interactions and Mobility Edges: Observing the Generalized
  Aubry-Andr\'e Model}},}\ }\href {\doibase 10.1103/PhysRevLett.126.040603}
  {\bibfield  {journal} {\bibinfo  {journal} {Phys. Rev. Lett.}\ }\textbf
  {\bibinfo {volume} {126}},\ \bibinfo {pages} {040603} (\bibinfo {year}
  {2021})}\BibitemShut {NoStop}%
\bibitem [{\citenamefont {Wang}\ \emph {et~al.}(2022)\citenamefont {Wang},
  \citenamefont {Zhang}, \citenamefont {Li}, \citenamefont {Wu}, \citenamefont
  {Liu}, \citenamefont {Mei}, \citenamefont {Hu}, \citenamefont {Xiao},
  \citenamefont {Ma}, \citenamefont {Chin},\ and\ \citenamefont
  {Jia}}]{Wang2022Observation}%
  \BibitemOpen
  \bibfield  {author} {\bibinfo {author} {\bibfnamefont {Yunfei}\ \bibnamefont
  {Wang}}, \bibinfo {author} {\bibfnamefont {Jia-Hui}\ \bibnamefont {Zhang}},
  \bibinfo {author} {\bibfnamefont {Yuqing}\ \bibnamefont {Li}}, \bibinfo
  {author} {\bibfnamefont {Jizhou}\ \bibnamefont {Wu}}, \bibinfo {author}
  {\bibfnamefont {Wenliang}\ \bibnamefont {Liu}}, \bibinfo {author}
  {\bibfnamefont {Feng}\ \bibnamefont {Mei}}, \bibinfo {author} {\bibfnamefont
  {Ying}\ \bibnamefont {Hu}}, \bibinfo {author} {\bibfnamefont {Liantuan}\
  \bibnamefont {Xiao}}, \bibinfo {author} {\bibfnamefont {Jie}\ \bibnamefont
  {Ma}}, \bibinfo {author} {\bibfnamefont {Cheng}\ \bibnamefont {Chin}}, \ and\
  \bibinfo {author} {\bibfnamefont {Suotang}\ \bibnamefont {Jia}},\ }\bibfield
  {title} {\enquote {\bibinfo {title} {{Observation of Interaction-Induced
  Mobility Edge in an Atomic Aubry-Andr\'e Wire}},}\ }\href {\doibase
  10.1103/PhysRevLett.129.103401} {\bibfield  {journal} {\bibinfo  {journal}
  {Phys. Rev. Lett.}\ }\textbf {\bibinfo {volume} {129}},\ \bibinfo {pages}
  {103401} (\bibinfo {year} {2022})}\BibitemShut {NoStop}%
\bibitem [{\citenamefont {White}\ \emph {et~al.}(2020)\citenamefont {White},
  \citenamefont {Haase}, \citenamefont {Brown}, \citenamefont {Hoogerland},
  \citenamefont {Najafabadi}, \citenamefont {Helm}, \citenamefont {Gies},
  \citenamefont {Schumayer},\ and\ \citenamefont
  {Hutchinson}}]{white2020observation}%
  \BibitemOpen
  \bibfield  {author} {\bibinfo {author} {\bibfnamefont {Donald~H.}\
  \bibnamefont {White}}, \bibinfo {author} {\bibfnamefont {Thomas~A.}\
  \bibnamefont {Haase}}, \bibinfo {author} {\bibfnamefont {Dylan~J.}\
  \bibnamefont {Brown}}, \bibinfo {author} {\bibfnamefont {Maarten~D.}\
  \bibnamefont {Hoogerland}}, \bibinfo {author} {\bibfnamefont {Mojdeh~S.}\
  \bibnamefont {Najafabadi}}, \bibinfo {author} {\bibfnamefont {John~L.}\
  \bibnamefont {Helm}}, \bibinfo {author} {\bibfnamefont {Christopher}\
  \bibnamefont {Gies}}, \bibinfo {author} {\bibfnamefont {Daniel}\ \bibnamefont
  {Schumayer}}, \ and\ \bibinfo {author} {\bibfnamefont {David A.~W.}\
  \bibnamefont {Hutchinson}},\ }\bibfield  {title} {\enquote {\bibinfo {title}
  {Observation of two-dimensional {Anderson} localisation of ultracold
  atoms},}\ }\href {\doibase 10.1038/s41467-020-18652-w} {\bibfield  {journal}
  {\bibinfo  {journal} {Nat Commun}\ }\textbf {\bibinfo {volume} {11}},\
  \bibinfo {pages} {4942} (\bibinfo {year} {2020})}\BibitemShut {NoStop}%
\bibitem [{\citenamefont {Dikopoltsev}\ \emph {et~al.}(2022)\citenamefont
  {Dikopoltsev}, \citenamefont {Weidemann}, \citenamefont {Kremer},
  \citenamefont {Steinfurth}, \citenamefont {Sheinfux}, \citenamefont
  {Szameit},\ and\ \citenamefont {Segev}}]{Dikopoltsev2022Observation}%
  \BibitemOpen
  \bibfield  {author} {\bibinfo {author} {\bibfnamefont {Alex}\ \bibnamefont
  {Dikopoltsev}}, \bibinfo {author} {\bibfnamefont {Sebastian}\ \bibnamefont
  {Weidemann}}, \bibinfo {author} {\bibfnamefont {Mark}\ \bibnamefont
  {Kremer}}, \bibinfo {author} {\bibfnamefont {Andrea}\ \bibnamefont
  {Steinfurth}}, \bibinfo {author} {\bibfnamefont {Hanan~Herzig}\ \bibnamefont
  {Sheinfux}}, \bibinfo {author} {\bibfnamefont {Alexander}\ \bibnamefont
  {Szameit}}, \ and\ \bibinfo {author} {\bibfnamefont {Mordechai}\ \bibnamefont
  {Segev}},\ }\bibfield  {title} {\enquote {\bibinfo {title} {Observation of
  anderson localization beyond the spectrum of the disorder},}\ }\href
  {\doibase 10.1126/sciadv.abn7769} {\bibfield  {journal} {\bibinfo  {journal}
  {Sci. Adv.}\ }\textbf {\bibinfo {volume} {8}},\ \bibinfo {pages} {7769}
  (\bibinfo {year} {2022})}\BibitemShut {NoStop}%
\bibitem [{\citenamefont {Roy}\ \emph {et~al.}(2018)\citenamefont {Roy},
  \citenamefont {Khaymovich}, \citenamefont {Das},\ and\ \citenamefont
  {Moessner}}]{Roy2018Multifractality}%
  \BibitemOpen
  \bibfield  {author} {\bibinfo {author} {\bibfnamefont {Sthitadhi}\
  \bibnamefont {Roy}}, \bibinfo {author} {\bibfnamefont {Ivan~M.}\ \bibnamefont
  {Khaymovich}}, \bibinfo {author} {\bibfnamefont {Arnab}\ \bibnamefont {Das}},
  \ and\ \bibinfo {author} {\bibfnamefont {Roderich}\ \bibnamefont
  {Moessner}},\ }\bibfield  {title} {\enquote {\bibinfo {title}
  {{Multifractality without fine-tuning in a Floquet quasiperiodic chain}},}\
  }\href {\doibase 10.21468/SciPostPhys.4.5.025} {\bibfield  {journal}
  {\bibinfo  {journal} {SciPost Phys.}\ }\textbf {\bibinfo {volume} {4}},\
  \bibinfo {pages} {025} (\bibinfo {year} {2018})}\BibitemShut {NoStop}%
\bibitem [{Lin()}]{Lin2023Model}%
  \BibitemOpen
  \href@noop {} {}\bibinfo {note} {Xiaoshui Lin, Ming Gong, and Guangcan Guo,
  In preparing.}\BibitemShut {Stop}%
\end{thebibliography}%

\end{document}